\documentclass[a4paper,11pt]{article} 
\pdfoutput=1 
\usepackage{jheppub} 

\usepackage{xcolor}                         
\usepackage{colortbl}
\usepackage[normalem]{ulem}                 
\usepackage{booktabs}                       
\usepackage{multirow,bigdelim}              

\usepackage{soul}                           
\usepackage{amsmath,amssymb,mathtools,mathrsfs}      
\usepackage{slashed}                        
\usepackage{amsthm}
\usepackage{bbm,dsfont}                     

\usepackage{bm}                             
\usepackage[vcentermath]{youngtab}          
\usepackage{ytableau}                       
\usepackage{suffix}	                        
\usepackage{lscape}	                        

\usepackage{graphicx}                       
\usepackage{subfig}	                        
\usepackage{tikz}                           
\usepackage{float}
\usepackage[export]{adjustbox}              

\usepackage{url}                            

\usepackage[numbers,sort&compress]{natbib}  

\usepackage[colorlinks=true,urlcolor=blue,anchorcolor=blue,citecolor=blue,filecolor=blue,linkcolor=blue,menucolor=blue,pagecolor=blue,linktocpage=true,pdfproducer=medialab,pdfa=true]{hyperref}	                    
\usepackage{ifthen}	                        

\numberwithin{equation}{section}        
\numberwithin{table}{section}
\numberwithin{figure}{section}

\graphicspath{{figs/}}

\newcommand{\re}{{\rm e}}
\newcommand{\ri}{{\mathsf{i}}}
\newcommand{\rd}{{\rm d}}

\newcommand{\nn}{\nonumber \\}

\newcommand{\und}[1]{\underline{#1}}

\def\mc{\mathcal}
\def\md{\mathbf}
\def\mf{\mathfrak}
\def\ms{\mathsf}
\def\mr{\mathscr}

\def\IC{\mathbb{C}}

\def\IN{\mathbb{N}}
\def\IP{\mathbb{P}}

\def\IR{\mathbb{R}}
\def\IZ{\mathbb{Z}}

\newcommand{\disc}{\text{disc}}
\newcommand{\wh}[1]{\widehat{#1}}

\newcommand{\CA}{\mc{A}}

\newcommand{\CV}{\mc{V}}

\newcommand{\ep}{\epsilon}

\newcommand{\pd}{\partial}

\def\bra#1{\left\langle #1 \right|}
\def\ket#1{\left| #1 \right\rangle}
\def\vev#1{\left\langle #1 \right\rangle}

\newcommand{\dDe}[1]{\hspace{-0.5ex}
  \raisebox{.23ex}{
  {$\stackrel{\raisebox{-.23ex}{$\scriptscriptstyle\bullet$}}\Delta_{
  \raisebox{-.23ex}[1ex][0ex]{$\scriptstyle#1$}}$}}
}

\newcommand{\dDes}[1]{\hspace{-0.5ex}
  \raisebox{.23ex}{
    {$\stackrel{\raisebox{-.23ex}{$\scriptscriptstyle\bullet$}}
      \Delta_{\raisebox{-.23ex}[1ex][0ex]{$\scriptstyle#1$}}^{\raisebox{-.23ex}[1ex][0ex]{$\scriptstyle(s)$}}$}}
}

\title{\boldmath Towards full instanton trans-series in Hofstadter's butterfly}%
\author{Jie Gu${}^a$, Zhaojie Xu${}^b$}%

\affiliation[a]{School of Physics and Shing-Tung Yau Center\\
  Southeast University, Nanjing 210096, China}%
\affiliation[b]{Department of Physics and Institute for Quantum Science and Technology,\\ Shanghai University,
	99 Shangda Road, Shanghai 200444, China}%

\emailAdd{jie-gu@seu.edu.cn}
\emailAdd{zeezj@shu.edu.cn}

\abstract{The trans-series completion of perturbative series of a wide
  class of quantum mechanical systems can be determined by combining
  the resurgence program with extra input coming from exact WKB
  analysis. In this paper, we reexamine the Harper-Hofstadter model
  and its spectrum, Hofstadter's butterfly in light of recent
  developments. We demonstrate the connection between the perturbative
  energy series of the Harper-Hofstadter model and the vev of
  $1/2$-BPS Wilson loop of 5d SYM and clarify the differences between
  their non-perturbative corrections. Taking insights from the cosine
  potential model, we construct the full energy trans-series for flux
  $\phi=2\pi/Q$ and provide numerical evidence with remarkably high
  precision. Finally, we revisit the problem of self-similarity of the
  butterfly and discuss the possibility of a completed version of the
  Rammal-Wilkinson formula. }


\begin{document}
\maketitle
\flushbottom

\noindent \textcolor{gray}{\caps{Version}: \today}

\section{Introduction}

The energy spectrum of the simple system of electrons on a
two-dimensional square lattice in a uniform magnetic field has a
surprisingly rich structure. After the early studies of Harper
\cite{Harper:1955}, in 1976 D.~Hofstadter pointed out
\cite{Hofstadter:1976zz} that this is a very peculiar system where the
electron spectrum has different features when the value of the
magnetic field is rational or irrational, and he derived a recursion
equation which allowed him to plot the energy spectrum of the electron
system against the magnetic field when the magnetic field is rational.
The resulting beautiful plot is later known as the Hofstadter's
butterfly due to its resemblance to a butterfly in shape, and it
raises many puzzling questions. Thanks to the periodicity of the
square lattice, the energy of the electron system displays a band
structure that depends on the Bloch angle. The Hofstadter's butterfly
indicates that the energy is a rather intricate function of both the
magnetic field and the Bloch angle, but the exact nature of such a
function is rather mysterious.  In addition, the Hofstadter's
butterfly has a fractal structure, which can be described by a
strong-weak field duality map
$(E,\phi) \rightarrow (\tilde{E},1/\phi)$, where $\phi$ is the
magnetic flux through a lattice plaquette, but the expression of the
mapped energy $\tilde{E}$ is yet unknown.  
Due to its simple-looking but rather intricate structure, the
Hofstadter's butterfly has also attracted attention of many physicists
and mathematicians
\cite{Krasovsky:2000rr,Marra:2023gio,Bellissard:1994lcg,Barelli:unpub},
and interesting connections to quantum integrable systems
\cite{Wiegmann:1994zz,Faddeev:1993uk}, quantum Hall effect
\cite{TKNN,Kohmoto:1983mit} and possibly high-temperature superconductivity
\cite{Hasegawa:1989ted} were discovered.
With the proof of the Ten-Martini problem \cite{Avila}, there are
still some unsolved mysteries for the Harper-Hofstadter model.


One way of studying the energy as a function of the magnetic field is
to consider the weak field limit, where the energy is treated as a
perturbative series in the magnetic field.  Such a perturbative
series, nevertheless, is oblivious to the Bloch angle and thus cannot
explain the band structure.  In fact, the rich band structure is known
to be caused by non-perturbative effects.  For instance, the bandwidth
is explained by the instanton effects in the path integral formalism
\cite{Freed:1990uw}.  However, fully understanding the
non-perturbative corrections including all-order instanton effects
would still be a challenge.

In recent years there have been several new developments that made the
solution of this problem a distinct possibility.  The first
development is the discovery of a surprising connection to an
unexpected territory. In 2016, Hatsuda, Katsura and Tachikawa found
\cite{Hatsuda:2016mdw} that the Harper-Hofstadter model is naturally related
to the 5d $\mc{N}=1$ $G=SU(2)$ Super Yang-Mills theory on
$S^1\times \IR^4$, or alternatively topological string theory on local
$\IP^1\times\IP^1$ as its string theory realization.  In the IR, the
5d gauge theory is completely characterized by an algebraic curve
called the Seiberg-Witten curve, and it was noticed that the curve
equation is the same as the Harper Hamiltonian without the magnetic
field. Turning on the magnetic field is equivalent to quantizing the
Seiberg-Witten curve. This allows us to calculate many quantities
efficiently in the Harper-Hofstadter model. For instance, the perturbative
energy series of the Harper-Hofstadter model is mapped to the perturbative
series of the Wilson loop, while the instanton corrections are
controlled by the free energy of the field theory, both of which can
be computed efficiently using the holomorphic anomaly equations
\cite{Bershadsky:1993cx,Bershadsky:1993ta,Huang:2010kf,Wang:2023zcb,Huang:2022hdo}.
As a result, the authors of \cite{Duan:2018dvj} were able to find the
complete one-instanton and the partial two-instanton corrections to
the energy series of the Harper-Hofstadter model.  This connection between
electrons in 2d lattices and supersymmetric field theory or string
theory was later extended to other models
\cite{Hatsuda:2017zwn,Hatsuda:2017dwx,Hatsuda:2020ocr}.

Another development is the powerful resurgence theory
\cite{Ecalle,Sauzin:2014intro,Marino:2012zq,Aniceto:2018bis}, which
claims that a perturbative series and its non-perturbative corrections
are intimately related, and that a subset of the non-perturbative
corrections can be extracted from the perturbative series itself.
Another result of \cite{Duan:2018dvj} was to use the resurgence
technique to confirm the (partial) two-instanton corrections to the
energy series in the Harper-Hofstadter model.  More importantly, in the 5d
SYM, the non-perturbative corrections to both the Wilson loops and the
free energies, at least the part accessible by the resurgence
techniques have been solved in their entirety \cite{Gu:2022fss},
which are conjectured to be controlled by the BPS spectrum of the 5d
SYM.  These results should be reinterpreted in the Harper-Hofstadter model.

Finally, the exact WKB method \cite{Voros1983}, which is the
traditional WKB method enhanced by resurgence techniques, has been
very useful in deriving exact quantization conditions for 1d
non-relativistic quantum mechanical models.  See
\cite{wilkinson:1984critical,wilkinson:1984example} for earlier
analysis of the Harper-Hofstadter model with the WKB method.  Recently, the
exact WKB method has been revisited \cite{vanSpaendonck:2023znn} so
that in many 1d QM models the full energy trans-series including
instanton corrections to all orders are written down, and they all
share the same universal structure.  It implies that even if we do not
know the exact quantization conditions of a 1d QM model a priori, we
can still try to construct the full energy trans-series by looking for
a family of well-organized basic building blocks fitting the
trans-series coefficients.

In this paper, we will combine the results of all these recent
developments to construct the full energy trans-series for the
Harper-Hofstadter model in some special cases.  The Harper-Hofstadter model is
equivalent to a 1d relativistic QM model.  We assume the universal
structure of the full energy trans-series is still valid.  We then
borrow elements from the 5d SYM to construct the basic building blocks
of this general structure.  We use resurgence results from the 5d SYM
to find a subset of the trans-series coefficients and use high
precision numerical calculation to find the remaining coefficients.
With this method, we are able to confirm that the universal structure
of the energy trans-series is still valid and find the full energy
trans-series when the magnetic flux is $\phi = 2\pi/Q$ for natural
number $Q$.  Taking the logic of \cite{vanSpaendonck:2023znn} in
reverse, we infer the exact quantization conditions from the full
energy trans-series and find that it is in some sense a ``double
copy'' of the exact quantization condition of the Mathieu equation,
i.e.~the 1d non-relativistic QM model with a cosine potential
\cite{Delabaere:1992sos,ZinnJustin:2004mer,Dunne:2014uwm,Sueishi:2021xti}.

In this process, we clarify a subtlety in the identification between
the non-perturbative corrections to the perturbative Wilson loop in
the field theory and the non-perturbative corrections to the
perturbative energy series of the Harper-Hofstadter model, which is akin to
the transition from the large $N$ expansion to the conventional series
discussed recently in \cite{Marino:2024yme}.

In addition, we also find that the energy trans-series is very
sensitive to the nature of the magnetic flux.  If the magnetic flux is
$\phi = 2\pi P/Q$ with coprime natural numbers $P,Q$ and $P>1$, the
trans-series coefficients change, and they display a peculiar feature
related to the strong-weak magnetic field duality, and hence could
shed some light on the fractal structure of the spectrum possibly.  We
also study the expansion of the energy around some rational values of the
magnetic flux, extending the Rammal-Wilkinson formula
\cite{wilkinson:1984critical,wilkinson:1984example}.

The remainder of the paper is organized as the following.  In
Sec.~\ref{sc:review}, we review the previous results of the
Harper-Hofstadter model, including the secular equation that computes
the energy spectrum exactly when the magnetic field is rational, and
the semi-classical analysis, including the instanton corrections from
the path integral formalism.  In Sec.~\ref{sc:resurgence}, we collect
results from recent developments, including a short introduction to
the resurgence ideas that we will need, the exact WKB method and its
implication for the energy trans-series, and the connection between
the Harper-Hofstadter model and the 5d SYM.  In Sec.~\ref{sc:trans},
using these results, we construct the full energy trans-series for the
Harper-Hofstadter model step by step for flux $\phi = 2\pi /Q$. In
Sec.~\ref{sc:Csb}, we make some attempts to characterize the splitting
bands for $P>1$. We revisit the self-similarity structure of the
butterfly and provide evidence for a possible exact version of the
Rammal-Wilkinson formula.
%
%
Finally we conclude and give a list of open problems in
Sec.~\ref{sc:con}.

\section{Hofstadter's butterfly and its semiclassical analysis}
\label{sc:review}

The Harper-Hofstadter problem concerns the movement of electrons in
the square lattice of ions with the presence of uniform magnetic
field. According to Bloch's theorem, this can be effectively captured
by a single electron wavefunction obeying the almost Mathieu equation,
which was first studied by Harper \cite{Harper:1955}.  We will quickly
review this story here.

\subsection{Harper-Hofstadter equation}
\label{sc:harper}

Let us consider an electron moving in a two dimensional plane with a
doubly periodic electric potential induced by a square lattice of
ions.  By the tight binding approximation, the Hamiltonian operator of
the electron is
\begin{equation}
  \ms{H} = 2\cos \frac{a}{\hbar}\ms{p}_x + 2\cos \frac{a}{\hbar}\ms{p}_y.
\end{equation}
Here $a$ is the lattice spacing, $\hbar$ the reduced Planck constant,
and $\ms{p}_x,\ms{p}_y$ are the two independent momentum operators in
the $x$- and $y$-directions, and they commute with each other.  This
Hamiltonian allows a single continuous band of energy in the range
\begin{equation}
  -4 \leq E \leq 4.
\end{equation}

If we impose in addition a uniform magnetic field of field strength
$B$ perpendicular to the plane, the Hamiltonian operator has to be
modified where we replace $\ms{p}_x,\ms{p}_y$ by the operators of
canonical momenta $\ms{\Pi}_x,\ms{\Pi}_y$
\begin{equation}
  \ms{H} = \re^{\frac{\ri a}{\hbar}\ms{\Pi}_x}
  + \re^{-\frac{\ri a}{\hbar}\ms{\Pi}_x}
  + \re^{\frac{\ri a}{\hbar}\ms{\Pi}_y}
  + \re^{-\frac{\ri a}{\hbar}\ms{\Pi}_y}.
\end{equation}
Here the canonical momenta are defined by
\begin{equation}
  \vec{\ms{\Pi}} = \vec{\ms{p}} + e \vec{A}
\end{equation}
and the two components no longer commute
\begin{equation}
  [\ms{\Pi}_x,\ms{\Pi}_y] = -\ri\hbar e (\pd_x A_y - \pd_y A_x) =
  -\ri\hbar e B.
\end{equation}
We will call this the Harper-Hofstadter model.

We can simplify the notation by defining the scaled operators
\begin{equation}
  \ms{x} = \frac{a}{\hbar}\ms{\Pi}_x,\quad \ms{y} = -\frac{a}{\hbar}\ms{\Pi}_y
\end{equation}
with the commutator
\begin{equation}
  [\ms{x},\ms{y}] = \frac{\ri a^2 e B}{\hbar} =: \ri\phi,
\end{equation}
so that the Hamiltonian simply reads
\begin{equation}
  \label{eq:hamiltonian}
  \ms{H} = \re^{\ri\ms{x}} + \re^{-\ri\ms{x}}
  + \re^{\ri\ms{y}} + \re^{-\ri\ms{y}}.
\end{equation}
This is equivalent to a relativistic one dimensional quantum
mechanical model where $\ms{x},\ms{y}$ play the roles of the position
and the momentum operators respectively, and the flux through a
lattice plaquette $\phi$ plays the role of the reduced Planck
constant.  In the position representation, the time-independent
Schr\"odinger equation reads
\begin{equation}
  \psi(x+\phi) + \psi(x-\phi) + 2\cos x \psi(x) = E\psi(x).
\end{equation}
Introduce the parametrization
\begin{equation}
  x = n\phi + \delta,\quad \psi_n(\delta) = \psi(n\phi+\delta),
\end{equation}
we arrive at the famous Harper's equation
\begin{equation}
  \label{eq:harper}
  \psi_{n+1} + \psi_{n-1} + 2\cos(n\phi+\delta) \psi_n = E\psi_n.
\end{equation}

\subsection{Butterfly at rational fluxes}
\label{sc:butterfly}

When the magnetic flux $\phi$ is rational of the form
\begin{equation}
  \phi = 2\pi\alpha = 2\pi\frac{P}{Q},\quad P,Q\in \IN,\; (P,Q) =1,
\end{equation}
the energy spectrum of the Harper's equation can be derived relatively
easily, as first found out by \cite{Hofstadter:1976zz}.  In this case,
the Harper's equation is invariant under the shift $n\rightarrow n+Q$,
and we can introduce the Bloch wavefunction
\begin{equation}
  \psi_n(\delta) = \re^{\ri k n}u_n(\delta,k),
\end{equation}
where $k$ is the Bloch wavenumber, and $u_n$ is periodic with
\begin{equation}
  u_{n+Q}(\delta,k) = u_n(\delta,k).
\end{equation}
The matrix of the Hamiltonian operator in the Hilbert space then
truncates to finite size and we have the eigenvalue equation
\begin{equation}
  H_Q \cdot u_Q = E u_Q,\quad u_Q = (u_0,u_1,\ldots,u_{Q-1})^T
\end{equation}
where $H_Q$ is the matrix
\begin{equation}
  H_Q(\delta,k) =
  \begin{pmatrix}
    2\cos\delta &\re^{\ri k} &  &  & \re^{-\ri k}\\
    \re^{-\ri k} & 2\cos(\delta + 2\pi \frac{P}{Q}) & \re^{\ri k} &&\\
    & \re^{-\ri k} & 2\cos(\delta+4\pi \frac{P}{Q}) & \re^{\ri k}&\\
    &&\ddots&\ddots&\\
    \re^{\ri k} &&&\re^{\ri k}&2\cos(\delta + 2\pi(Q-1)\frac{P}{Q})
  \end{pmatrix}.
\end{equation}
The energy spectrum is solved from the secular equation
\begin{equation}
  \label{eq:secular0}
  F_{P/Q}(E,\delta,k):= \det(H_Q - E\md{1}_Q) = 0.
\end{equation}
The left hand side defines a degree $Q$ polynomial in $E$, which we
denote by $F_{P/Q}(E,\delta,k)$, and it indicates that for fixed
$\delta,k$, there are $Q$ eigen-energies.

It can be shown (see e.g.~\cite{Hasegawa:1990}) that the secular
equation can be equivalently written as
\begin{equation}
  \label{eq:secular1}
  F_{P/Q}(E,0,0) = 2(\cos Qk + \cos Q\delta) =: 2(\cos\theta_x+\cos\theta_y).
\end{equation}
Here we have denoted $Qk,Q\delta$ respectively by $\theta_x,\theta_y$.
They can be treated on equal footing: both of them are periodic with
$\theta_{x,y}\rightarrow \theta_{x,y}+2\pi$, and the secular equation
is not changed by exchanging $\theta_x,\theta_y$.  In fact, it was
pointed out in \cite{Duan:2018dvj} that the Harper-Hofstadter model is
special in the sense that when the flux $\phi$ is rational there can
be Bloch angles in both the $x$- and $y$-directions, and
$\theta_x,\theta_y$ defined here are precisely these two Bloch angles.

\begin{figure}
	\centering
	\includegraphics[width=12cm]{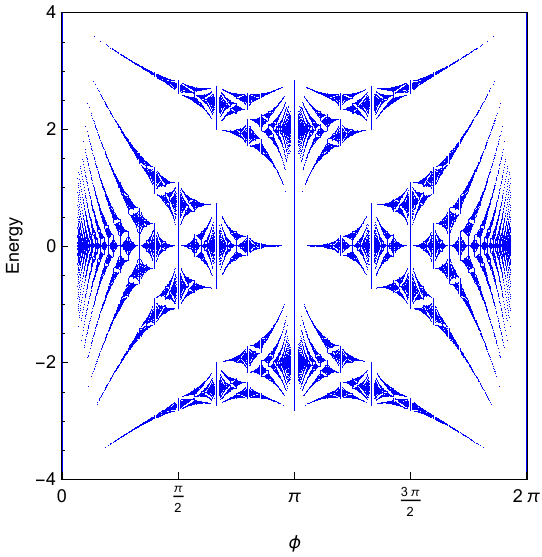}
	\caption{Hofstadter. We plot the band structure for $P/Q$ with
          $(P,Q)=1$ and $Q$ up to 60.}
	\label{Hofbutterfly}
\end{figure}

The secular equation \eqref{eq:secular1} also indicates that by
varying $\theta_x,\theta_y$ in their respective domain
, the $Q$ eigen-energies are broadened to $Q$ continuous energy bands,
where the top and the bottom edges correspond respectively
to 
$\theta_x=\theta_y = 0$ and $\theta_x=\theta_y = \pi$.  The spectrum
of energy as a function of the flux is plotted in
Fig.~\ref{Hofbutterfly}\footnote{There's also a useful open-source package \cite{HofstadterTools} that can be used to compute the band structure of a generalized Hofstadter model on any regular Euclidean lattice, as well as its key properties.}.  As the Harper's equation \eqref{eq:harper},
and therefore the energy $E$, is invariant under the shift
$\phi\rightarrow \phi+2\pi$, the plot is restricted to the domain of
$\phi \in [0,2\pi]$.

This plot of spectrum in Fig.~\ref{Hofbutterfly} is the famous
Hofstadter's butterfly.  It has a striking fractal structure, which
implies that the energy spectrum as a function of the flux has very
rich non-perturbative structures, which we try to understand.

\subsection{Semi-classical analysis}
\label{sc:classical}

As mentioned in Sec.~\ref{sc:harper}, the Harper's equation can be
viewed as the Schr\"odinger equation of a relativistic one dimensional
quantum mechanical model with $\phi$ plays the role of the reduced
Planck constant.  It is natural then to treat the spectrum problem
semiclassically, and consider the energy $E$ first as a perturbative
series in $\phi$.

In one-dimensional non-relativistic quantum mechanical problems, in
principle the perturbative energy series can be calculated by the
Rayleigh-Schr\"odinger perturbation theory, but in practise, one
cannot go very far.  Instead, it is more efficient to use the method
of Bender and Wu \cite{Bender:1969si,Bender:1990pd}, which makes the
ansatz that the wavefunction is a deformation of that of the harmonic
oscillator, and which allows very fast calculation of the perturbative
energy around any local minimum of a polynomial potential where the
second derivative of the potential does not vanish.  This algorithm
was made into a \texttt{Mathematica} package called \texttt{BenderWu}
in \cite{Sulejmanpasic:2016fwr}, which was expanded in
\cite{Gu:2017ppx} to allow relativistic systems whose Hamiltonians are
polynomials of $\re^{\pm x},\re^{\pm y},x,y$.  As pointed out in
\cite{Duan:2018dvj}, after a Wick rotation
$x,y\rightarrow \ri x,\ri y$, our Hamiltonian \eqref{eq:hamiltonian}
falls into this category.  With the help of the \texttt{BenderWu}
package, one can easily calculate the perturbative energy series for
the Harper-Hofstadter model up to close to 100 terms, and the first
few terms are
\begin{equation}
  \label{eq:Epert}
  E^{\text{pert}}(\nu,\phi) = 4-2\nu\phi + (\frac{1}{16}+\frac{\nu^2}{4})\phi^2
  + (-\frac{\nu}{128}-\frac{\nu^3}{96})\phi^3 + \ldots
\end{equation}
with the Landau level $\nu = N + 1/2$, $N = 0,1,2,\ldots$.

As discussed at the end of Sec.~\ref{sc:butterfly}, the spectrum of
the Harper-Hofstadter model has significant non-perturbative
corrections, which presumably come from instanton effects.  The
leading instanton corrections can be computed by the path integral
formalism \cite{Duan:2018dvj}.
For this purpose, we recall that in  the more standard one dimensional
periodic 
quantum mechanical model with a single Bloch angle $\theta$, we can define
the Bloch wavefunction
\begin{equation}
  \psi_\theta(x) = \sum_{N\in\IZ} \re^{-\ri N\theta} \psi_0(x+Na)
\end{equation}
where $\psi_0(x)$ is the approximate eigenstate wavefunction centered
around the origin, as well as the twisted thermal partition function
\begin{equation}
  \bra{\psi_{\theta'}}\re^{-HT/\phi}\ket{\psi_\theta}  =
  2\pi\delta(\theta-\theta')Z_\theta(T) =
  2\pi\delta(\theta-\theta')\sum_{\nu\in \IN+1/2}\re^{-E_\theta(\nu)T/\phi}.
\end{equation}
Assuming the energy spectrum is not degenerate, the ground state
energy can then be computed using the twisted thermal partition
function
\begin{equation}
  E_{\theta}(1/2) = -\lim_{T\rightarrow \infty}\frac{\phi}{T}\log Z_\theta(T).
\end{equation}
The twisted partition function can be computed in path integral, and
in the semiclassical limit with $\phi\rightarrow 0$, it decomposes by
\begin{equation}
  Z_\theta = Z_\theta^{(0)} + Z_\theta^{(+1)}+ Z_\theta^{(-1)} + Z_\theta^{(+2)}+ Z_\theta^{(-2)} +\ldots
\end{equation}
where
\begin{equation}
  Z_\theta^{(n)} = \int [\mc{D}x]_{n} \re^{-S_E/\phi +\ri
    n\theta}, 
\end{equation}
with the boundary conditions
\begin{equation}
  [\mc{D}x]_n: \;x(-T/2)=0,\; x(+T/2)=na,
\end{equation}
which describe precisely an $|n|$-instanton configuration.

In the case of the Harper-Hofstadter model, this path integral
analysis was performed in \cite{Duan:2018dvj} for the cases of
$\phi = 2\pi/Q$, i.e.~$P=1$.  It was observed that one can find
1-instanton configurations in both the $x$- and $y$-directions, with
the corresponding Bloch angles $\theta_x,\theta_y$, and the instanton
action is identically
\begin{equation}
  S_c = 8C
\end{equation}
$C$ being the Catalan's number.  With instantons in both $x$- and
$y$-directions, and in both positive and negative directions, the
one-instanton correction of the ground state energy
is\footnote{There is a typo in \cite{Duan:2018dvj}, where
  1-instanton amplitude should be increased by a factor of two.\label{ft:typo}}
\begin{equation}
  E^{(1)}_{\theta_x,\theta_y}(1/2,\phi) = 16(\cos\theta_x+\cos\theta_y)
  \left(\frac{\phi}{2\pi}\right)^{1/2}\re^{-S_c/\phi}(1+\ldots).
\end{equation}
This can be checked by comparing with the bandwidth of each energy
band
\begin{equation}
  \mathrm{bw}_N(\phi)\approx \big|E^{(1)}_{0,0}(\nu,\phi) - E^{(1)}_{\pi,\pi}(\nu,\phi)\big|,
\end{equation}
which at the leading order is controlled by the 1-instanton correction.
In fact, with this method, one finds numerically that at any Landau
level \cite{Duan:2018dvj}\footnote{See footnote \ref{ft:typo}.}
\begin{equation}
  \label{eq:E1}
  E^{(1)}_{\theta_x,\theta_y}(\nu,\phi) =
  (\cos\theta_x+\cos\theta_y)(-1)^N\frac{16\cdot  8^{N}}{\pi^N N!}
  \left(\frac{\phi}{2\pi}\right)^{1/2-N}\re^{-S_c/\phi}(1+\ldots).
\end{equation}

If we take into account instanton corrections of all orders, we expect
the non-perturbative energy series to be of the form
\begin{equation}
  \label{eq:E-trans0}
  E(\nu,\phi) = E^{(0)}(\nu,\phi) + \sum_{n\geq 1} E^{(n)}_{\theta_{x,y}}(\nu,\phi),
\end{equation}
where the $n$-instanton correction is of the order
\begin{equation}
  E^{(n)} \sim \re^{-n S_c/\phi}.
\end{equation}
We will make this expression more concrete in Sec.~\ref{sc:trans}.

\section{Resurgence, exact WKB, and 5d SYM}
\label{sc:resurgence}

\subsection{Resurgence program}
\label{sc:res-program}

We give here a quick overview of the resurgence theory.  See
\cite{Ecalle,Sauzin:2014intro,Marino:2012zq,Aniceto:2018bis} for more
detailed discussion.  Given a perturbative series $\varphi(z)$, which
is of 1-Gevrey type, meaning that its coefficients grow factorially
fast
\begin{equation}
  \varphi(z) = \sum_{n=0}^\infty a_n z^n,\quad a_n \sim n!
\end{equation}
such that it has zero radius of convergence, there is a well-studied
procedure called Borel resummation to evaluate, or to resum such a
divergent series.

For this purpose, we first construct the Borel transform of the
1-Gevrey series
\begin{equation}
  \wh{\varphi}(\zeta) = \sum_{n=0}^\infty \frac{a_n}{n!}\zeta^n,
\end{equation}
which is regular in the neighborhood of the origin.  It can be
analytically continued to the entire complex plane, and let us make
the mild assumption that it has a discrete set $\Omega\subset\IC$ of
singular points, known as the Borel singularities.

Let us define Stokes lines in the complex $z$-plane, which are rays
from the origin and whose inclinations are the arguments of the Borel
singularities.  These Stokes lines divide the complex $z$-plane into
disjoint cones.  For any value of $z$ inside a cone, we can define the
Borel resummation
\begin{equation}
  \mr{S}\varphi(z) = \frac{1}{z}\int_0^{\re^{\ri\arg
      z}\infty}\re^{-\zeta/z}\wh{\varphi}(\zeta)\rd\zeta. 
\end{equation}
If $z$ is on a Stokes line, naive definition above of Borel
resummation would fail as the integration contour will be obstructed
by a Borel singularity.  In this case, we have to define not one but a
pair of lateral Borel resummations by slightly raising or lowering the
inclination of the integration contour to bypass the Borel singularity
\begin{equation}
  \mr{S}^{(\pm)}\varphi(z) = \frac{1}{z}
  \int_0^{\re^{\ri(\arg z\pm\epsilon)}\infty}
  \re^{-\zeta/z}\wh{\varphi}(\zeta)\rd\zeta,
\end{equation}
and the two resummations differ by an exponentially suppressed
discrepancy known as the Stokes discontinuity
\begin{equation}
  \text{disc}_{\theta}\varphi(z) = \mr{S}^{(+)}\varphi(z)
  -\mr{S}^{(-)}\varphi(z)\sim \re^{-1/z},
\end{equation}
where $\theta$ is the inclination of the Stokes line.

Suppose there is a sequence of Borel singularities
${kA} = A,2A,3A,\ldots$ which share the same argument as $z$ and which
obstruct the naive integration contour for the Borel resummation.
According to the resurgence theory, the Stokes discontinuity can be
attributed in a precise manner to these Borel singularities.
In fact, each such Borel singularity $kA$ represents a non-trivial
saddle point in the theory, and to each is associated a new 1-Gevrey
series $\varphi^{(k)}$, such that
\begin{equation}
  \disc_\theta \varphi(z) = \sum_{k=1}^\infty
  \ms{S}_{k}\re^{-kA/z}\mr{S}^{(-)}\varphi^{(k)}(z). 
\end{equation}
The proportionality constants $\ms{S}_{k}$ are known as the Borel
residues and they depend on the normalization of the series
$\varphi^{(k)}$.

It is sometimes more useful to encode in a different manner
contributions of individual singular points to the Stokes
discontinuity.  For instance, we can introduce a map of power series
known as Stokes automorphism
\begin{equation}
  \mf{S}_\theta\varphi(z): = \varphi(z)
  + \sum_{k=1}^\infty  \ms{S}_{k}\re^{-kA/z}\varphi^{(k)}(z).
\end{equation}
so that
\begin{equation}
  \mr{S}^{(+)}\varphi(z) = \mr{S}^{(-)}\mf{S}_\theta \varphi(z).
\end{equation}
which has the property that it is an automorphism in the ring of power
series.  Alternatively, we can introduce pointed alien derivatives
associated to each of the Borel singularities
\cite{Ecalle}\footnote{Alien derivatives can also be introduced even
  if the Borel singularities are spaced unevenly.}
\begin{align}
  \mf{S}_\theta \varphi(z) =
  &\exp\left(\sum_{k=1}^\infty
    \dDe{kA}\right)\varphi(z)\nn =
  &\varphi(z)+\dDe{A}\varphi(z) +
    \left(\dDe{2A}+\frac{1}{2}(\dDe{A})^2\right)\varphi(z) + \ldots
\end{align}
Each alien derivative is a map of 1-Gevrey power series, and in
particular, we have
\begin{equation}
  \dDe{kA} \varphi(z) = \mc{S}_{k}\re^{-kA/z}\varphi^{(k)}(z)
\end{equation}
The coefficients $\mc{S}_{k}$ here are called the Stokes constants,
and they are related to the Borel residues by simple combinatoric
formulas.
The alien derivatives have very nice properties: they follow the
Leibniz rule and chain rule, just like ordinary derivatives, and
furthermore commute with ordinary derivations. 

As the new series $\varphi^{(k)}$ uncovered from the original
perturbative series $\varphi$ are also 1-Gevrey, the same resurgence
analysis of Borel singularities and Stokes discontinuities can be
repeated, revealing even more Borel singularities and the associated
additional 1-Gevrey power series.  Together all these 1-Gevrey series
are said to form a \emph{minimal resurgent structure} starting from
$\varphi$ \cite{Gu:2021ize}.

From the discussion of resurgent structure, a paradigm to study
generic perturbative series called resurgence program can be
formulated.  One distinguishes between the weak resurgence program and
the strong resurgence program \cite{DiPietro:2021yxb}.  The \emph{weak
  resurgence program} conjectures that any physical quantity that
allows a perturbative expansion $\varphi$ can be expressed in terms of
the Borel resummation of a trans-series, whose leading contribution is
the perturbative series $\varphi$.  Trans-series is a rather broad
concept, see \cite{Edgar:2008usf} for a good exposition.  The most
common form of trans-series, which is enough for us, is
\begin{equation}
  \Phi(z) = \varphi(z) + \sum_k c_k \re^{-A_k/z} \varphi^{(A_k)}(z)
\end{equation}
where $\varphi^{(A_k)}(z)$ are usually power series just like
$\varphi(z)$, but may also contain terms with $\log(z)$.  The
\emph{strong resurgence program} in addition requires that all
$\varphi^{(A_k)}$ belong to the minimal resurgence structure starting
from $\varphi$.  In many scenarios, the strong resurgence program is
too strong, and only a subset of $\varphi^{(A_k)}$, which is sometimes
called the \emph{minimal trans-series} \cite{vanSpaendonck:2023znn},
belong to the minimal resurgent structure.  Regardless, the
trans-series coefficients $c_k$ associated to this subset of power
series will jump as we cross a Stokes line in order to compensate for
the Stokes discontinuity so that the exact physical quantity can be a
continuous function of $z$ and is ambiguity free.
In general, as $z$ moves in the complex plane, crossing various
Stokes lines, all ingredients of the minimal resurgent structure will
appear in the full trans-series.

\subsection{Structure of trans-series from exact WKB}
\label{sc:WKB}

Following the weak resurgence program, the exact energy eigenvalue
should be the Borel resummation of an energy trans-series.
In 1d QM models, a particularly powerful method to derive such an
energy trans-series is to solve exact quantization conditions (EQCs)
obtained via the exact WKB method
\cite{Voros1983
}, which is based on the resurgence theory.  It was implied in
\cite{vanSpaendonck:2023znn} that full energy trans-series seems to
have a universal structure, which we explain.
In later sections, we will demonstrate that the full energy
trans-series of the Harper-Hofstadter model shares this universal
structure.

Suppose we have the Schr\"odinger equation for a 1d non-relativistic
QM model
\begin{equation}
  H(\ms{x},\ms{y}) \psi(x) = E\psi(x),
\end{equation}
which is a second order ODE.  We can write down the WKB ansatz for the
wavefunction
\begin{equation}
  \psi(x) = \exp\left(\frac{\ri}{\hbar}\int_*^x P(x',\hbar)\rd x'\right).
\end{equation}
Here $P(x,\hbar)$ is a formal power series
\begin{equation}
  P(x,\hbar) = \sum_{n=0}^\infty P_n(x)\hbar^n.
\end{equation}
The coefficients $P_n(x)$ can be solved by plugging in the WKB ansatz
into the Schr\"odinger equation.  The leading coefficient
$P_0(x) = \pm y(x)$ is the momentum satisfying the classical equation
\begin{equation}
  \label{eq:HE}
  H(x,y) = E.
\end{equation}
Higher order coefficients $P_{n\geq 1}(x)$ can be solved recursively.

\begin{figure}
  \centering%
  \includegraphics[height=4.5cm]{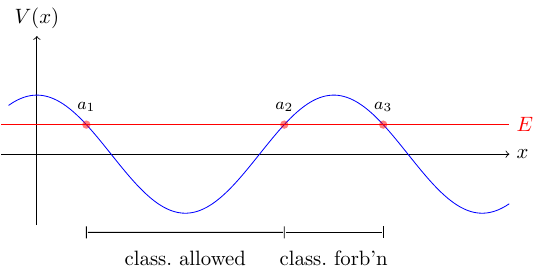}
  \caption{Classically allowed and forbidden regions}
  \label{fig:classical}
\end{figure}

If we promote $x,y$ to complex variables, the classical equation
\eqref{eq:HE} defines a complex curve known as the WKB curve $\Sigma$.
We will assume that the WKB curve is of genus one, so that it has two
independent 1-cycles, called the A-cycle $\gamma_A$ and the B-cycle
$\gamma_B$ with intersection number $\vev{\gamma_A,\gamma_B} = 1$.  We
choose the A-cycle $\gamma_A$ and B-cycle $\gamma_B$ so that when
projected to the complex $x$-plane, they are mapped to respectively
the classical allowed and classically forbidden regions,
cf.~Fig.~\ref{fig:classical}.  We can then define the quantum A- and
B-periods
\begin{subequations}
  \begin{align}
    t(E,\hbar) =
    &\frac{1}{2\pi}\sum_{n=0}^\infty
      \hbar^{2n}\oint_{\gamma_A}P_{2n}(x)\rd x,\label{eq:tE}\\
    t_D(E,\hbar) =
    &-\ri \sum_{n=0}^\infty \hbar^{2n}\oint_{\gamma_B}P_{2n}(x)\rd x.\label{eq:tDE}
  \end{align}
\end{subequations}
They are also known as the perturbative and the non-perturbative
quantum periods, as they are responsible for respectively the
perturbative and non-perturbative contributions to the quantization
conditions that we will see momentarily.  And the normalization in
\eqref{eq:tE} and \eqref{eq:tDE} are chosen so that they are positive
in the leading order.  Both quantum periods are power series in
$\hbar$, and the leading terms are classical periods of the 1-form
$\lambda = y(x)\rd x$.  In addition, both quantum periods are
1-Gevrey, as
\begin{equation}
  \oint_\gamma P_{2n}(x)\rd x \sim (2n)!
\end{equation}
The difference is that when $\hbar >0$, the perturbative quantum
period is not Borel summable, so that a prescription of lateral
resummation is needed, while the non-perturbative quantum period is
Borel summable, and a vanilla version of Borel resummation is
applicable.

In general, the EQCs for the eigen-energy $E$ take the form
\begin{equation}
  \label{eq:EQC}
  1 + \CV_A =  f(\CV_B^{1/2},\CV_A^{1/2}),
\end{equation}
with the Voros symbols
\begin{equation}
  \CV_A = \re^{2\pi\ri t(E,\hbar)/\hbar},
  \quad \CV_B = \re^{-t_D(E,\hbar)/\hbar}.
\end{equation}
Here $f(u,v)$ is certain single-valued function of $u,v$, and it
vanishes in the $u\rightarrow 0$ limit, corresponding to the
semi-classical limit $\hbar\rightarrow 0$.  For instance this is true
for the cubic mode, the double-well model (see
e.g.~\cite{vanSpaendonck:2023znn}), and in particular for the cosine
model, whose Schr\"odinger equation is the famous Mathieu equation.
This is the one dimensional quantum mechanical model with the
Hamiltonian
\begin{equation}
  H = \frac{y^2}{2} +1 - \cos(x),
\end{equation}
and it can be regarded as the non-relativistic limit of the
Hamiltonian \eqref{eq:hamiltonian} of the Harper-Hofstadter model.
The EQC for this model is well-known and it reads  
\cite{Delabaere:1992sos,ZinnJustin:2004mer,Dunne:2014uwm,Sueishi:2021xti}
\begin{equation}
  \label{eq:EQC-cosine}
  D^{\pm}_\theta = 1 + \CV_A^{\mp 1}(1+\CV_B) -
  2\sqrt{\CV_A^{\mp 1}\CV_B}\cos\theta = 0.
\end{equation}
Depending on the choice of the lateral resummation $\mr{S}^{\pm}$ of
the perturbative quantum period, one of the two quantization
conditions $D^{\pm}_\theta$ is used.  Here $\theta$ is the Bloch
angle.

To solve the energy trans-series, we first consider the semi-classical
limit $\hbar\rightarrow 0$, with the non-perturbative contributions
due to $\re^{-t_D/\hbar}$ turned off.  The EQC is reduced to
\begin{equation}
  1 + \CV_A = 0 
\end{equation}
which is equivalent to the all-orders Bohr-Sommerfeld quantization
conditions
\begin{equation}
  \label{eq:PQC}
  t(E,\hbar) = \hbar\nu,
\end{equation}
where $\nu = N+1/2$, $N=0,1,2,\ldots$.  Let $\mc{E}(t,\hbar)$ be the
inverse of $t(E,\hbar)$ as a function of $E$, the perturbative energy
series is
\begin{equation}
  E^{(0)}(\nu,\hbar) = \mc{E}(t=\hbar\nu,\hbar).
\end{equation}

To solve the EQC \eqref{eq:EQC} with the non-perturbative corrections
turned on, one can assume that $t$ is a small deviation from
$\hbar\nu$
\begin{equation}
  t = \hbar (\nu + \Delta \nu),
\end{equation}
and solve $\Delta t$ from the equation
\begin{equation}
  \label{eq:imeq}
  1 - \re^{2\pi\ri\Delta \nu} =
  f(\re^{-\frac{1}{2\hbar}t_D(\nu+\Delta \nu,\hbar)},
  \pm\ri\,\re^{\pi\ri\Delta \nu}),
\end{equation}
where the exponent $t_D(\nu,\hbar)$ is
\begin{equation}
  t_D(\nu,\hbar) := t_D(E=E^{(0)}(\nu,\hbar),\hbar),
\end{equation}
while the full enery trans-series is then obtained by substituting the
deformed $\nu+\Delta\nu$ for $\nu$ in the perturbative series
\begin{equation}
  \label{eq:E-E0Del}
  E(\nu,\hbar) =e^{\Delta\nu\partial\nu}E^{(0)}(\nu,\hbar)= E^{(0)}(\nu+\Delta\nu,\hbar).
\end{equation}
This is the strategy pursued in \cite{vanSpaendonck:2023znn}, where it
is proposed to recast the eq.~\eqref{eq:imeq} in the form
of\footnote{Note that the power series on the right hand side must
  start with $k=1$ as it should vanish in the semi-classical limit
  $\hbar\rightarrow 0$.}
\begin{equation}
  \label{eq:DeltR}
  \Delta \nu = R(\lambda(\nu+\Delta\nu)) = \sum_{k=1}^\infty r_k \lambda(\nu+\Delta\nu)^k,\quad
  \lambda(\nu) = \re^{-\frac{1}{2\hbar}t_D(\nu,\hbar)},
\end{equation}
whose solution as a trans-series can be explicitly written down via
the Lagrange inversion theorem (see e.g.~\cite{Gessel:2016lai, Surya2023}).  One
then finds that the full energy trans-series obtained via
\eqref{eq:E-E0Del} has the general structure
\cite{vanSpaendonck:2023znn}
\begin{equation}
  \label{eq:E-trans-gen}
  E(\nu,\hbar) 
  = E^{(0)}(\nu,\hbar) + \sum_{n=1}^\infty
  \sum_{m=0}^{n-1} u_{n,m}(\underline{r})E^{(n,m)}(\nu,\hbar).
\end{equation}
The basic building blocks are the basic trans-series
\begin{equation}
  \label{eq:Enm}
  E^{(n,m)}(\nu,\hbar) =
  \left(\frac{\pd}{\pd\nu}\right)^m\left(\frac{\pd
      E^{(0)}(\nu,\hbar)}{\pd \nu}\re^{-t_D(\nu,\hbar)/\hbar}\right).
\end{equation}
All of $E^{(n,m)}$ with $m=0,1,2,\ldots,n-1$ account for the
$n$-instanton corrections.  Note that $E^{(n,m)}$ with
$n\geq 2, m\geq 1$ may contain $\log(\hbar)$ terms and they arise due
to instanton / anti-instanton interactions.
The trans-series coefficients read
\begin{equation}
  \label{eq:uB}
  u_{n,m}(\underline{r}) = \frac{1}{n!}B_{n,m+1}(1!r_1,2!r_2,\ldots,(n-m)!r_{n-m}),
\end{equation}
where $B_{n,m+1}$ are the incomplete Bell's polynomials.  Finally the
weak resurgence program requires that the Borel resummation of the
full energy trans-series gives the exact value of energy 
in the regime $\hbar\nu \ll 1$.

Compared to the general full trans-series \eqref{eq:E-trans0}, the
structure \eqref{eq:E-trans-gen} is much simpler.  The basic building
blocks $E^{(n,m)}$ only depend on two ingredients, the perturbative
energy series $E^{(0)}(\nu,\hbar)$ and the non-perturbative quantum
period $t_D(\nu,\hbar)$, and once they are identified, the remaining
job is to fix relatively simpler trans-series coefficients $u_{n,m}$.

For non-relativistic one dimensional quantum mechanical models whose
Schr\"odinger equations are second order difference equations, the
exact WKB method is still applicable, although it is difficult to
write down EQCs in this way as the connection formulas are yet not
competely clear (see \cite{DelMonte:2024dcr} though for recent
progress).  The EQCs have been written down in some examples by other
methods \cite{Grassi:2014zfa,Wang2015,Hatsuda:2015qzx,Franco:2015rnr},
for instance via the TS/ST corresondence
\cite{Grassi:2014zfa,Codesido:2015dia}, but these EQCs have a more
complicated form.
Nevertheless, we will see in later sections that the universal
structure \eqref{eq:E-trans-gen} for energy trans-series also holds
for the Harper-Hofstadter model, as least when $\phi = 2\pi/Q$.
Furthermore, the basic building blocks can be readily written down.
For instance, it has already been shown \cite{Duan:2018dvj} via
examples of 1-instanton corrections that the non-perturbative quantum
period can be easily computed as it has an interesting interpretation
in supersymmetric field theories, which we quickly review.

\subsection{5d SYM and its resurgent structure}
\label{sc:5d-SYM}

We will be interested in 5d $\mc{N}=1$ supersymmetric Yang-Mills
theory with gauge group $G = SU(2)$ on $S^1\times \IR^4$.  The IR
effective theory is described by the Seiberg-Witten curve given by the
equation \cite{Katz:1996fh,Klemm:1996bj}\footnote{We have chosen the
  special so-called diagonal slice in the moduli space where the
  radius of $S^1$ is one.}
\begin{equation}
  \Sigma: \quad \re^x + \re^{-x} + \re^{y} + \re^{-y} - u = 0
\end{equation}
The Seiberg-Witten curve is equipped with the meromorphic 1-form
\begin{equation}
  \lambda = y\rd x,
\end{equation}
and its integration along closed 1-cycles on $\Sigma$ are known as
classical periods.

A 5d $\mc{N}=1$ supersymmetric theory usually has degenerate vacuum
states, and they form a moduli space $\mc{M}$.  Due to $\mc{N}=1$
supersymmetry, the moduli space has the structure of a special
K\"ahler manifold, which means that in any patch of the moduli space,
one can choose a basis of flat coordinates to locally parametrise the
moduli space, and these flat coordinates are paired with their
conjugates (see e.g.~\cite{klemm2018b})
\begin{equation}
  \left(t^a,\frac{\pd F_0}{\pd t^a}\right),\quad
  a=1,\ldots,\frac{1}{2}\dim\mc{M},
\end{equation}
so that they are related to each other via a single function called
the prepotential $F_0$.  Such a choice of flat coordinates is called
choosing a frame.  Here both $t^a$ and $\pd_{t^a}F_0$ are integral
periods of the meromorphic form $\lambda$ over the Seiberg-Witten
curve, and togethe with $4\pi^2 \ri$ they span the period lattice.

In the case of SYM, the moduli space is $\IP^1$, parametrised by
$z = 1/u^2$, and when quantum corrections are taken into account, it
has three singular points, located at $z = 0,1/16,\infty$, known as
the large radius point, the conifold point, and the orbifold point.
The neighborhood of the conifold point will be of particular interest
for us.  Here, the suitable flat coordinate and its conjugate are (see
e.g.~\cite{Marino:2015ixa})
\begin{subequations}
  \begin{align}
    t_c =
    &\frac{1}{\pi}\left(\frac{1}{\pi} G^{3,2}_{3,3}\left(
    \begin{smallmatrix}
      \tfrac{1}{2},\tfrac{1}{2};1\\
      0,0,0
    \end{smallmatrix};16z \right) - \pi^2\right),\label{eq:tc}\\
    \frac{\pd F_0(t_c)}{\pd t_c} =
    &-\pi\Big(\log(z) +
      4z{}_4F_3(1,1,\tfrac{3}{2},\tfrac{3}{2};2,2,2;16z)\Big)
      +\pi\ri t_c,\label{eq:tcD}
  \end{align}
\end{subequations}
which have the property that $t_c(z=1/16) = 0$.
We use a slightly different convention of $\pd_{t_c}F_0(t_c)$ than in
the literature, as we will be interested in the regime $z > 1/16$,
where our convention has the property that
$\pd_{t_c}F_0(t_c)\in \IR_+$.

We couple the gauge theory to a background gravity by turning on the
Omega background \cite{Nekrasov:2002qd} and restrict ourselves to the
so-called Nekrasov-Shatashvili (NS) limit \cite{Nekrasov:2009rc}.  The
Seiberg-Witten curve is promoted to a quantum operator known as the
quantum Seiberg-Witten curve \cite{Aganagic:2011mi}.  It is a
relativistic Schr\"odinger equation
\begin{equation}
  \label{eq:H-SYM}
  \ms{H}^{\text{SYM}}\psi = u\psi
\end{equation}
with the Hamiltonian operator
\begin{equation}
  \ms{H}^{\text{SYM}} = \re^{\ms{x}} + \re^{-\ms{x}} + \re^{\ms{y}} + \re^{-\ms{y}},
\end{equation}
and $\ms{x},\ms{y}$ satisfy the canonical quantization condition
\begin{equation}
  [\ms{x}, \ms{y}] = \ri\hbar.
\end{equation}
As shown in \cite{Hatsuda:2015qzx}, the quantum Seiberg-Witten curve
can be identified with the Hamiltonian of the two particle closed
relativisitc Toda lattice \cite{Ruijsenaars1989RelativisticTS}.  On
the other hand, if we make the Wick rotation \cite{Hatsuda:2016mdw}
\begin{equation}
  (x,y)\rightarrow (\ri x,\ri y),
\end{equation}
as well as the map
\begin{equation}
  \label{eq:hbar-phi}
  \hbar \rightarrow -\phi,
\end{equation}
the quantum Seiberg-Witten curve \eqref{eq:H-SYM} can be identified
with the Hamiltonian operator \eqref{eq:hamiltonian} of the
Harper-Hofstadter model.  While the EQCs of the relativistic Toda
lattice have been written down, with no distinction between the
rational and irrational $\hbar$ \cite{Grassi:2014zfa,Wang2015}, those
for the Harper-Hofstadter model are much more complicated.  This is
akin to the difference between the Mathieu equation and the modified
Mathieu equation.

We will be interested in finding in the Harper-Hofstadter model the
full energy trans-series and the implied EQCs in the form of
\eqref{eq:EQC}.  We fix our convention and define the perturbative
quantum period and non-perturbative quantum periods \eqref{eq:tE},
\eqref{eq:tDE} so that their leading terms are respectively $t_c$ and
$\pd_{t_c}F_0(t_c)$ in \eqref{eq:tc},
\eqref{eq:tcD}.


Two interesting observables can be defined in the 5d SYM.  The first
is the vev $W_{\md{r}}(t,\hbar)$ of the half-BPS Wilson loop operator
in the fundamental representation $\md{r} = \square$ where the Wilson
loop wraps the $S^1$ and is located at the center of $\IR^4$.  In the
NS limit, the perturbative Wilson loop vev is a power series in
$\hbar$ \cite{Huang:2022hdo,Wang:2023zcb}
\begin{equation}
  \label{eq:W}
  W_\square(t,\hbar) = \sum_{n=0}^\infty W_n(t)\hbar^{2n}.
\end{equation}
where the coefficients $W_n(t)$ are functions over the moduli space.
More importantly it is identified with the perturbative eigenvalue of
the Hamiltonian operator $\ms{H}^{\text{SYM}}$
\cite{Nekrasov:2009rc,Gaiotto:2014ina,Bullimore:2014awa}
\begin{equation}
  u = W_\square(t,\hbar),
\end{equation}
and therefore also with the perturbative energy series of the
Harper-Hofstadter model with the dictionary \eqref{eq:hbar-phi}.  In
the semi-classical limit with $\hbar = -\phi \rightarrow 0$, the
energy of the Harper-Hofstadter model is
$4$, 
corresponding to the conifold point singularity at $z = 1/u^2 =  1/16$.  We
should thus evaluate the Wilson loop vev in the conifold frame.  The
first few coefficients are
\begin{subequations}
  \begin{align}
    W_0(t_c) = &4+2t_c + \frac{t_c^2}{4} + \ldots,\\
    W_1(t_c) = &\frac{1}{16} + \frac{t_c}{128} + \frac{3t_c^2}{1024} + \ldots,\\
    W_2(t_c) = &\frac{13}{24576} -\frac{151t_c}{393216} +
                 \frac{159t_c^2}{524288} + \ldots.
  \end{align}
\end{subequations}
It is easy to see that indeed \eqref{eq:W} reproduces
\eqref{eq:Epert} through
\begin{equation}
  \label{eq:E-W}
  E^{(0)}(\nu,\phi)  = W_\square(t_c,\hbar)\Big|_{t_c = -\phi\nu,\hbar = -\phi}.
\end{equation}
%

Another interesting physical observable is the NS free energy, which
is also a perturbative power series in $\hbar$ \cite{Nekrasov:2009rc}
\begin{equation}
  \label{eq:FNS}
  F_{\text{NS}}(t_c,\hbar) = \sum_{n=0}^\infty F_n(t_c)\hbar^{2n}.
\end{equation}
In the conifold frame, the perturbative free energy can be decomposed
in terms
\begin{equation}
  \label{eq:Fn}
  F_{n}(t_c) = F^{\text{sing}}(t_c) + F^{\text{reg}}(t_c)
\end{equation}
where the singular parts are
\begin{subequations}
  \label{eq:Fsing}
  \begin{align}
    &F^{\text{sing}}_0(t_c) = \frac{t_c^2}{2}\left(\log\left(-\frac{t_c}{16}\right)-\frac{3}{2}\right),\\ 
    &F^{\text{sing}}_1(t_c) = -\frac{1}{24}\log \left(-\frac{t_c}{16^2}\right),\\
    &F^{\text{sing}}_n(t_c) =
    \frac{
      (1-2^{1-2n})B_{2n}}{(2n)(2n-1)(2n-2)t_c^{2n-2}},\quad
    n\geq 2.
  \end{align}
\end{subequations}
while the first few terms of the regular parts are
\begin{subequations}
  \label{eq:Freg}
  \begin{align}
    &F^{\text{reg}}_0(t_c) = -8C t_c-\frac{t_c^3}{48}+ \frac{5t_c^4}{4608} - \frac{7t_c^5}{61440}+\ldots \\
    &F^{\text{reg}}_1(t_c) = -\frac{11t_c}{192} + \frac{49t_c^2}{9216}
      - \frac{77t_c^3}{73728}+\ldots\\
    &F^{\text{reg}}_2(t_c) = -\frac{101}{221184} -
      \frac{889t_c}{2949120} + \frac{181981t_c^2}{707788800} + \ldots.
  \end{align}
\end{subequations}
It was found out by calculations in the 1-instanton sector that the
non-perturbative quantum period $t_D$ can be identified with the free
energy through \cite{Duan:2018dvj}
\begin{equation}
  \label{eq:tD-F}
  t_D(\nu,\phi) = \frac{\pd}{\pd t_c}F(t_c,\hbar)\Big|_{t_c =
    -\phi\nu,\hbar = -\phi}.
\end{equation}
%


These identifications between quantities in the Harper-Hofstadter
model and observables in 5d SYM is very useful, as both the
perturbative Wilson loop vev and the perturbative free energy can be
computed very efficiently through the holomorphic anomaly equations
\cite{Bershadsky:1993cx,Krefl:2010fm,Huang:2010kf,Huang:2022hdo,Wang:2023zcb}.
More importantly, both of them turn out to be 1-Gevrey divergent power
series, and their resurgent structures have been recently completely
understood \cite{Gu:2022fss}. 

\begin{figure}
  \centering
  \subfloat[$z = 1/32$]{\includegraphics[height=5cm]{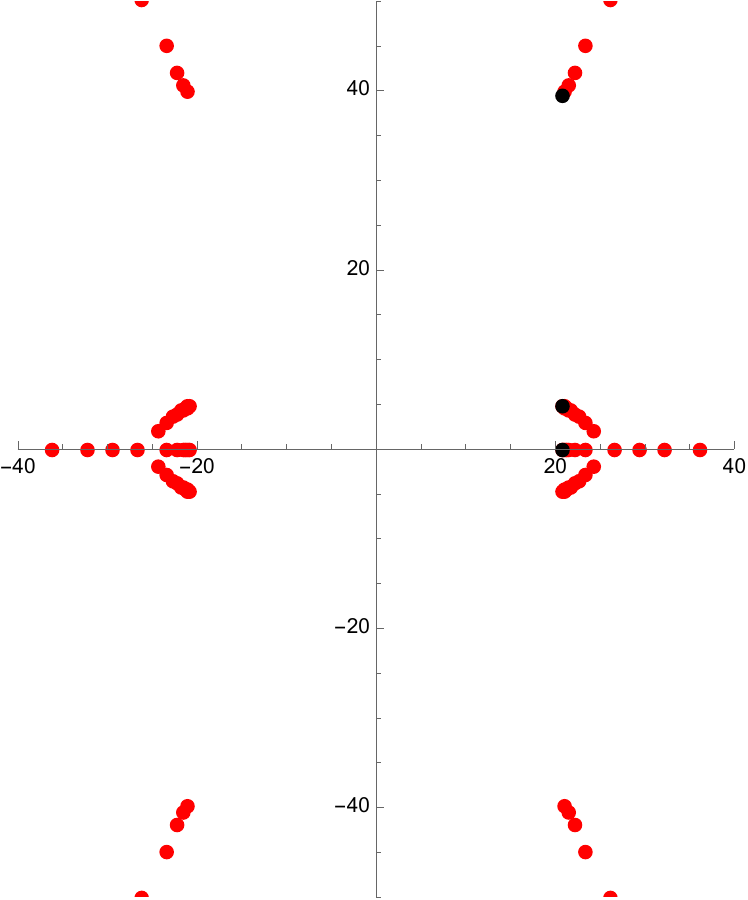}\label{fig:W-brl-sz}}\hspace{6ex}
  \subfloat[$z = 3/32$]{\includegraphics[height=5cm]{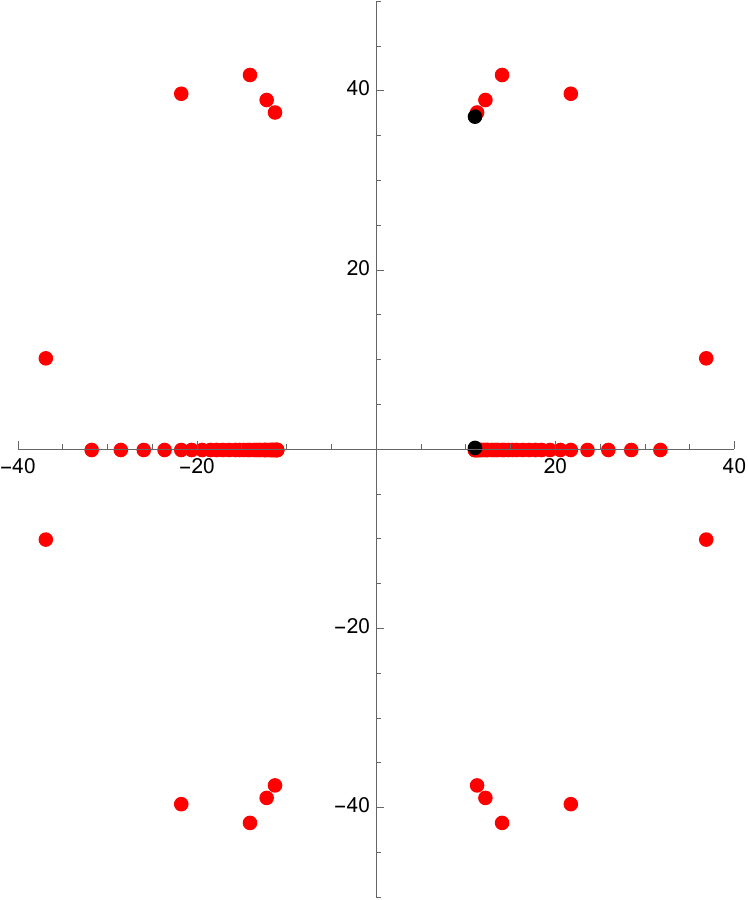}\label{fig:W-brl-lz}}
  \caption{Borel singularities for Wilson loop vev in the 5d SYM
    theory with (a) $z<1/16$ and (b) $z>1/16$ respectively.  In the
    left figure with $z<1/16$, the Borel singularities marked by black
    dots have charge vectors $\gamma = (2,-1,0)$ (on the real axis),
    $(2,0,0)$ (slightly away), $(2,-1,1)$ (far off in the first
    quadrant).  In the right figure with $z>1/16$, the Borel
    singularities marked by black dots have charge vectors
    $\gamma = (2,0,0)$ (on the positive real axis), $(2,1,1)$ (far off
    in the first quadrant).}
  \label{fig:W-brl}
\end{figure}

First of all, each Borel singularity corresponds conjecturally to a
BPS state of the 5d SYM.  The position of the Borel singularity is a
classical period\footnote{We use a slightly different convention from
  \cite{Gu:2022fss}.}
\begin{equation}
  A_\gamma = p \pd_{t_c}F_0(t_c) + 2\pi\ri q t_c + 4\pi^2 \ri r,
  \quad \gamma =  (p,q,r),
\end{equation}
which is the central charge of the BPS state, and the lattice charge
$\gamma$ is the electromagnetic charge of the BPS state.  For free
energies, all the BPS states are conjectured to appear, while for
Wilson loop vevs, only those whose charges have non-zero Dirac pairing
with the charge vector of the flat coordinate appear.  In the conifold
frame where the flat coordinate is $t_c$, this means those BPS states
with $p\neq 0$.  We give examples of plots of Borel singularities for
Wilson loop vevs for $z$ on the real axis smaller than and greater
than the conifold point $1/16$ respectively in Figs.~\ref{fig:W-brl}.
These two plots indicate that in the case of $z < 1/16$, the BPS
states with small central charges are \footnote{The charge vectors
  differ from the usual convention in the literatuer by $(0,-1,0)$, as
  we shifted the definition of $\pd_{t_c}F_0(t_c)$.}
\begin{equation}
  \gamma = \pm (2,-1,0),\; \pm (2,0,0),\; \pm (2,-1,1).
\end{equation}
while in the case of $z > 1/16$, the  BPS states with small central
charges are
\begin{equation}
  \gamma = \pm (2,0,0), \; \pm (2,1,1).
\end{equation}
The difference of the BPS spectrum in different chambers of the moduli
space is known as the wall-crossing phenomenon \cite{Kontsevich2008,
  Gaiotto2009, Cecotti2009}, and here it is clearly demonstrated via
the change of Borel singularities of Wilson loop vevs.  See
\cite{Marino:2024yme} for additional demonstrations via the change of
Borel singularities of free energy.  \footnote{To be precise, Borel
  singularities of self-dual free energy are considered in
  \cite{Marino:2024yme}, but they should be one-to-one correspondent
  with Borel singularitie of NS free energy \cite{Gu:2023wum}.}

Secondly, the Stokes discontinuities across Stokes rays are best
illustrated through the alien derivatives.  If these are a sequence of
Borel singularities $A,2A,\ldots$, the alien derivatives associated to
these singular points are
\begin{subequations}
  \begin{align}
    \dDe{\ell A_{\gamma}} W(t_c,\hbar) =
    &\frac{S^{\text{BPS}}_{\gamma}}{2\pi\ri}\hbar\frac{(-1)^{\ell}}{\ell}
      p \pd_{t_c}W(t_c,\hbar)
      \re^{-\ell p\pd_t F_{\text{NS}}^{\sharp}(t,\hbar)/\hbar},\label{eq:DelW}\\
    \dDe{\ell A_{\gamma}}F_{\text{NS}}(t_c,\hbar) =
    &\frac{S^{\text{BPS}}_{\gamma}}{2\pi\ri}\hbar^2\frac{(-1)^{\ell-1}}{\ell^2}
      \re^{-\ell p\pd_{t_c} F_{\text{NS}}^\sharp(t_c,\hbar)/\hbar},\label{eq:DelF}
  \end{align}
\end{subequations}
where the superscript $\sharp$ means the leading term of free energy is
shifted
\begin{equation}
  F_0(t_c) \rightarrow F_0^\sharp(t_c) = F_0(t_c ) + \frac{\pi\ri
    q}{p}t_c^2 + \frac{4\pi^2\ri r}{p} t_c,
\end{equation}
so that
\begin{equation}
  A_\gamma = p\pd_{t_c} F_0^\sharp(t_c).
\end{equation}
Most importantly, the Stokes constant $S^{\text{BPS}}_\gamma$ is
conjectured to coincide with the multiplicity $\Omega_\gamma$ of the
BPS state with charge vector $\gamma$.  For Borel singularities of the
perturbative Wilson loop vev at $z < 1/16$ in Fig.~\ref{fig:W-brl-sz},
which corresponds to the weak coupling regime of the 5d SYM, the
Stokes constant of the singularity on the real axis with
$\gamma=(2,-1,0)$ and that of the singularity slightly away with
$\gamma=(2,0,0)$ and $\gamma=(2,-2,0)$ are respectively \cite{Marino:2024yme}
\begin{equation}
  \label{eq:S-weak}
  S^{\text{BPS,[weak]}}_{(2,-1,0)} = -4,\quad
  S^{\text{BPS,[weak]}}_{(2,0,0)} = S^{\text{BPS,[weak]}}_{(2,-2,0)} = 2,
\end{equation}
while for the Borel singularities at $z>1/16$ in
Fig.~\ref{fig:W-brl-lz}, which corresponds to the strong coupling
regime of the 5d SYM, the Stokes constant of the singularity on the
positive real axis with $\gamma = (2,0,0)$ is \cite{Marino:2024yme}
\begin{equation}
  \label{eq:S-strong}
  S^{\text{BPS,[strong]}}_{(2,0,0)} = 2.
\end{equation}

\section{Full trans-series of Hofstadter's butterfly}
\label{sc:trans}

In this section, following the weak resurgence program, we will
demonstrate that for the Harper-Hofstadter model, at least in the case
of $\phi = 2\pi/Q$, the exact energy spectrum is the Borel resummation
of the full energy trans-series with different Landau levels.  We
assume that the universal structure of the full trans-series
\eqref{eq:E-trans-gen} inspired from the analysis of the exact WKB
method, which we copy below,
\begin{equation}
  \label{eq:E-trans-gen1}
  E(\nu,\phi) = E^{(0)}(\nu,\phi) + \sum_{n=1}^\infty
  \sum_{m=0}^{n-1} u_{n,m}E^{(n,m)}(\nu,\phi),
\end{equation}
where we have replaced $\hbar$ by $\phi$, and in addition to the
perturbative series $E^{(0)}(\nu,\phi)$, the basic building blocks are
the basic trans-series
\begin{equation}
  E^{(n,m)}(\nu,\phi) =
  \left(\frac{\pd}{\pd\nu}\right)^m\left(\frac{\pd
      E^{(0)}(\nu,\phi)}{\pd \nu}\re^{-t_D(\nu,\phi)/\phi}\right).
\end{equation}
We have also discussed in Sec.~\ref{sc:5d-SYM} that
$E^{(0)}(\nu,\phi)$ and $t_D(\nu,\phi)$ can be identified with
perturbative Wilson loop vev and perturbative free energy from 5d
$SU(2)$ SYM via \eqref{eq:E-W} and \eqref{eq:tD-F}.  We will justify
the assumption \eqref{eq:E-trans-gen1} by calculating the trans-series
coefficients $u_{n,m}$ and then by making precision comparison with
the exact spectrum from the secular equation \eqref{eq:secular1}.

\subsection{Borel summability of perturbative energy series}
\label{sc:summability}


We first discuss the Borel summability of the perturbative energy
series, i.e. whether we can perform the vanilla version of the Borel
resummation or lateral resummations are needed.

\begin{figure}
  \centering
  \subfloat[$N=0$]{\includegraphics[height=4.5cm]{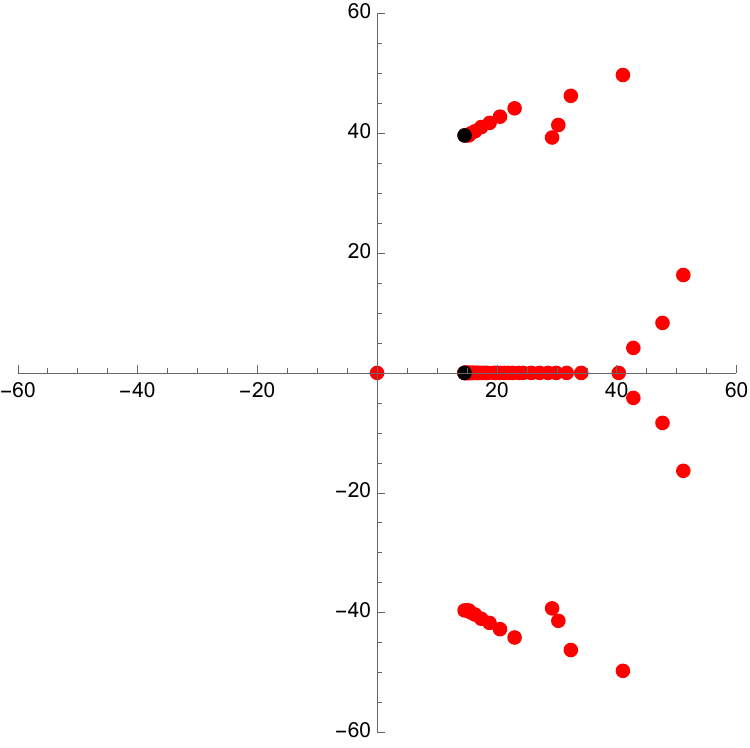}}\hspace{1ex}
  \subfloat[$N=1$]{\includegraphics[height=4.5cm]{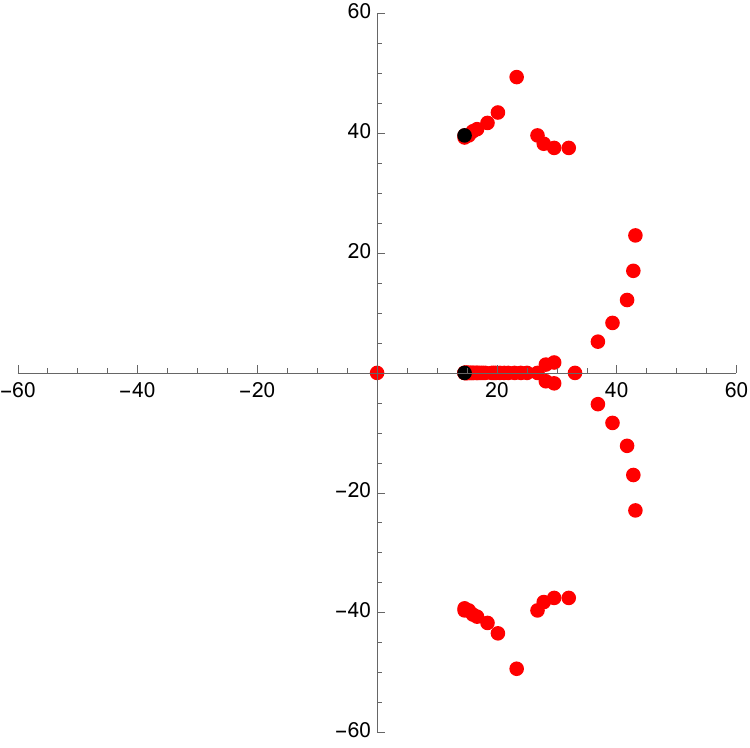}}\hspace{1ex}
  \subfloat[$N=2$]{\includegraphics[height=4.5cm]{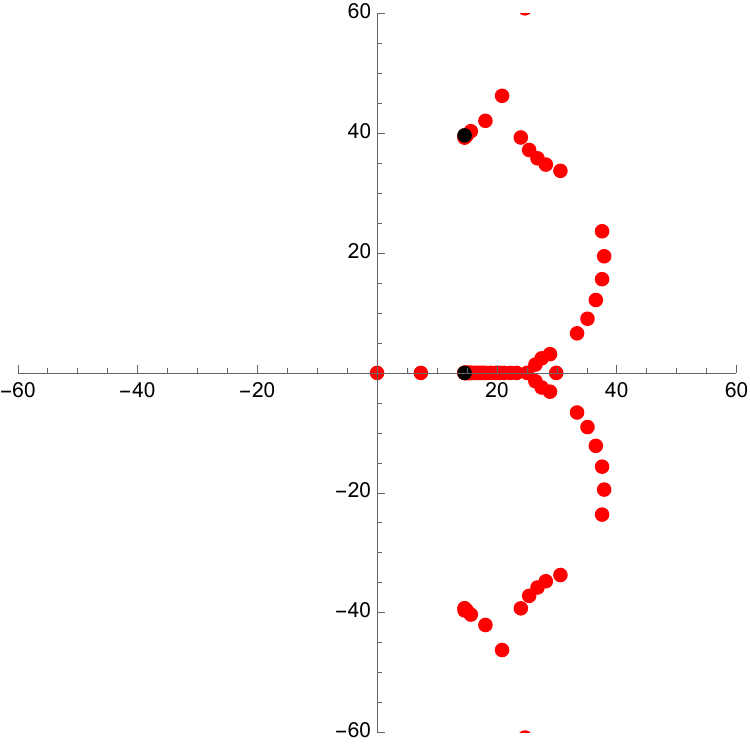}}\hspace{1ex}
  \caption{Borel singularities of perturbative energy series at Landau
    levels $N=0,1,2$.  The singularities marked by black dots on the
    positive real axis and off in the first quadrant in all three
    plots are $16C$ and $16C+4\pi^2\ri$, where $C$ is the Catalan
    number.  The arcs of singular points on the right periphery of
    each plot are due to numerical instability and thus are spurious.}
  \label{fig:E-brl}
\end{figure}

We collect about 200 terms of the perturbative energy series
$E^{(0)}(\nu,\phi)$ computed via either the \texttt{BenderWu} package
or via Wilson loop vev, the position of Borel singularities for the
energy series at different Landau levels are given in
Fig.~\ref{fig:E-brl}.
In all these plots, the dominant Borel singularities are
\begin{equation}
  16C,\quad 16C \pm 4\pi^2 \ri,
\end{equation}
%
where $C$ is the Catalan number.
There are several consequences of this pattern of Borel singularities.

The first consequence is that since the singular point $16C$ is on the
positive real axis, the naive version of Borel resummation fails.  We
have to adopt either choice of lateral Borel resummations, and the
ambiguity thus entailed should be compensated by an appropriate jump
of the trans-series coefficients.  We should update the structure of
full trans-series \eqref{eq:E-trans-gen1} to
\begin{equation}
  \label{eq:E-trans-gen2}
  E_{\theta_{x,y},\epsilon}(\nu,\phi) = E^{(0)}(\nu,\phi) + \sum_{n=1}^\infty
  \sum_{m=0}^{n-1} u_{n,m}(\theta_{x,y},\epsilon) E^{(n,m)}(\nu,\phi),
\end{equation}
such that the exact energy spectrum is
\begin{equation}
  E^{\text{ex}}_{\theta_{x,y}}(\nu,\phi) = \mr{S}^{\pm}
  (E_{\theta_{x,y}},\pm 1)(\nu,\phi).
\end{equation}
Note that only the trans-series coefficients $u_{n,m}$ depend on the
contour of lateral resummation as well as the Bloch angles, while the
trans-series building blocks $E^{(n,m)}$ do not.

Furthermore, the position of the most dominant Borel singularity
coincides with twice the 1-instantion $S_c = 8C$ discussed in
Sec.~\ref{sc:classical}.  This means that the 1-instanton contribution
$E^{(1,0)}$ is invisible from the Borel transform of the perturbative
series, and cannot be extracted from the latter by the resurgence
technique.  In other words, the strong resurgence program fails for
the Harper-Hofstadter energy spectrum, which is the second consequence
of the pattern of Borel singularities. This failure
of the strong resurgence program is due to the presence of Bloch
angles, and similar phenomena are already well-known, for instance in
the non-relativistic cosine model \cite{Dunne:2014uwm}.\footnote{Or
  even in simpler models like the double-well model, where the parity
  is a discrete analogue of the Bloch angle.}

Finally, we comment that even though the perturbative energy series of
the Harper-Hofstadter model and the perturbative Wilson loop vev of
the 5d SYM are identified via \eqref{eq:E-W}, the distribution of
their Borel singularities look rather different,
cf.~Figs.~\ref{fig:E-brl} and \ref{fig:W-brl}.  The most conspicuous
discrepancy is that the singular points in Figs.~\ref{fig:W-brl} for
the Wilson loop are left-right symmetric while those in
Figs.~\ref{fig:E-brl} for the Harper-Hofstadter energy are one-sided.
This can be explained by the observation that the perturbative Wilson
loop vev \eqref{eq:W} has the symmetry
\begin{equation}
  W(t_c,-\hbar) = W(t_c,\hbar),
\end{equation}
which is broken in the perturbative Harper-Hofstadter energy.
More detailed explanation is the following.

In the identification \eqref{eq:E-W} we use the dictionary
\begin{equation}
  \label{eq:N-lim}
  t_c = -\phi\nu,\quad \hbar = -\phi,
\end{equation}
 which means in the limit
$\hbar = -\phi\rightarrow 0$, both $t_c$ and $\hbar$ are sent to zero
simultaneously.  The relation between the series $E(\nu,\phi)$ and the
series $W(t_c,\hbar)$ is akin to the relation between a $1/N$
expansion and its 't Hooft limit, and the relation between their
respective resurgent structures, including the Borel singularities,
the non-perturbative series, and the Stokes constants, is recently
discussed in detail in \cite{Marino:2024yme}.  We will follow their
discussion and explain the relation between the resurgent structure of
$E(\nu,\phi)$ and $W(t_c,\hbar)$.

When we map from the resurgent structure of $W(t_c,\hbar)$ to that of
$E(\nu,\phi)$, several changes will happen.  The first change is that
half of the non-perturbative series will vanish and thus the
associated Borel singularities will disappear.  Let us examines how
this happens by deriving carefully the non-perturbative series for
$E(\nu,\phi)$ from those for $W(t_c,\hbar)$.  Recall that if there is
a sequence of Borel singularities $A_\gamma,2A_\gamma,\ldots$ for
$W(t_c,\hbar)$, the alien derivatives for these singular points are
given in \eqref{eq:DelW}, from which we read off the lowest
non-perturbative series, the contribution at $\ell=1$ excluding the
Stokes constant 
\begin{equation}
  W^{(1)}(t_c,\hbar) = \hbar\, p
  \frac{\pd W^{(0)}(t_c,\hbar)}{\pd t_c}
  \exp\left(-\frac{p}{\hbar}\frac{\pd F_{\text{NS}}^{\sharp}(t_c,\hbar)}{\pd t_c}\right).
\end{equation}
Here $p$ is the magnetic charge of the BPS state associated to the
Borel singularity $A_\gamma$, and we will assume that it is positive.

As seen in \eqref{eq:FNS}, \eqref{eq:Fn}, the free energy consists of
both the singular part and the regular part, and we discuss their
behavior after the dictionary \eqref{eq:N-lim} is applied.  The
coefficients of the regular part \eqref{eq:Freg} have the form
\begin{equation}
  F^{\text{reg}}_n(t_c) = \sum_{m\geq 1}f_{n,m}t_c^m,
\end{equation}
where we ignore the constant term.  After applying the dictionary
\eqref{eq:N-lim}, its derivative becomes\footnote{The $n=1$ terms
  vanishes because it only involves $f_{0,2}$, which according to
  \eqref{eq:Freg} is zero.}
\begin{equation}
  \label{eq:Freg-lim}
  \frac{\pd F^{\text{reg}}(t_c,\hbar)}{\pd t_c}
  \;\rightarrow\;
  f_{0,1} + \sum_{n\geq 2}(-\phi)^n \sum_{m=0}^{[n/2]}(n+1-2m)f_{m,n+1-2m}\nu^{n-2m},
\end{equation}
where the leading constant is, cf.~\eqref{eq:Freg}
\begin{equation}
  f_{0,1} = \frac{\pd F_0^{\text{reg}}(t_c)}{\pd t_c}\Big|_{t_c\rightarrow 0} = -8C.
\end{equation}

Next, we consider the singular part.  With \eqref{eq:Fsing}, one
finds, after applying the dictionary \eqref{eq:N-lim},
\begin{align}
  -\frac{1}{\hbar}\frac{\pd F^{\text{sing}}(t_c,\hbar)}{\pd t_c}
  =
  \nu -
  \nu\log\left(\frac{\nu}{16}\right)+\frac{1}{24\nu}
  +\sum_{n\geq
  2}\frac{1-2^{1-2n}}{(2n)(2n-1)}\frac{B_{2n}}{\nu^{2n-1}}-\nu\log\phi.
\end{align}
It is proposed in \cite{Codesido:2017jwp} that this can be regularised
as
\begin{equation}
  \label{eq:Fsing-lim}
  -\frac{1}{\hbar}\frac{\pd F^{\text{sing}}(t_c,\hbar)}{\pd t_c} \rightarrow
  \log\left(\frac{\sqrt{2\pi}\,16^\nu}{\Gamma(\nu+1/2)}\right) -
  \nu\log\phi,
\end{equation}
the reason being that the large $\nu$ expansion of the right hand side
reproduces the power series in $\hbar$ in the left hand side.  The
similar idea is used in \cite{Marino:2024yme}.

Combining \eqref{eq:Freg-lim} and \eqref{eq:Fsing-lim},
non-perturbative correction $W^{(1)}(t_c,\hbar)$ becomes 
\begin{align}
  W^{(1)}(t_c,\hbar)\;\rightarrow\;
  &p\frac{\pd E^{(0)}(\nu,\phi)}{\pd \nu}
    \left(\frac{\sqrt{2\pi}}{\Gamma(\nu+1/2)}\right)^p\left(\frac{16}{\phi}\right)^{p\nu}
    \re^{- \frac{p \mc{A}_{(p,r)}}{\phi} - 2\pi\ri q \nu}\nn
  &\exp \left(
    p\sum_{n\geq 2}(-\phi)^{n-1}\sum_{m=0}^{[n/2]}(n+1-2m)f_{m,n+1-2m}\nu^{n-2m}
    \right),
    \label{eq:W-inst}
\end{align}
where
\begin{equation}
  \mc{A}_{(p,r)} = -pf_{0,1}  - 4\pi^2 \ri r.
\end{equation}

The Wilson loop vevs have Borel singularities not only at
$A_\gamma,2A_\gamma,\ldots$, but also at the opposite sites
$-A_\gamma,-2A_{\gamma},\ldots$, and the associated non-perturbative
corrections are denoted by $W^{(0|\ell)}$.  The non-perturbative
corrections to the Harper-Hofstadter energy series inherited from
$W^{(0|\ell)}$, according to \cite{Marino:2024yme}, are obtained from
\eqref{eq:W-inst} (in the $\ell=1$ case) via
\begin{equation}
  (\phi,\nu) \; \rightarrow (-\phi,-\nu),
\end{equation}
which can be justified by the following symmetry of the perturbative
energy series
\begin{equation}
  E^{(0)}(-\nu,-\phi) = E^{(0)}(\nu,\phi).
\end{equation}
Therefore, for instance, the leading one of these non-perturbative
corrections is 
\begin{align}
  W^{(0|1)}(t_c,\hbar)\;\rightarrow\;
  &-\frac{p}{2\pi\ri}\frac{\pd E^{(0)}(-\nu,-\phi)}{\pd \nu}
    \left(\frac{\sqrt{2\pi}}{\Gamma(-\nu+1/2)}\right)^p\left(-\frac{16}{\phi}\right)^{p\nu}
    \re^{+ \frac{p \mc{A}_{p,r}}{\phi} - 2\pi\ri q \nu}\nn
  &\exp \left(
    p\sum_{n\geq 2}(+\phi)^{n-1}\sum_{m=0}^{[n/2]}(n+1-2m)f_{m,n+1-2m}(-\nu)^{n-2m}
    \right),
\end{align}
which vanishes for $\nu = 1/2,3/2,\ldots$ due to the pole of the Gamma
function in the denominator.  This explains why we do not see these
Borel singularities at opposite sites for the perturbative energy
series of the Harper-Hofstadter model.

A corollary of this analysis and \eqref{eq:W-inst} is that the set of
Borel singularities $A_{(p,q,r)}$ of the perturbative Wilson loop vev
with fixed $p,r$ but different $q$ all collapse to a single Borel
singularity $\mc{A}_{(p,r)}$ of the perturbative Harper-Hofstadter
energy series.  In addition, the alien derivative at this Borel
singularity is
\begin{equation}
  \label{eq:DelprE0}
  \dDe{\mc{A}_{(p,r)}}E^{(0)}(\nu,\phi) =
  \frac{S_{(p,r)}}{2\pi\ri}\mc{E}^{(p,r)}(\nu,\phi),
\end{equation}
where
\begin{align}
  \mc{E}^{(p,r)}(\nu,\phi)=
  &\frac{\pd E^{(0)}(\nu,\phi)}{\pd \nu}
    \left(\frac{\sqrt{2\pi}}{\Gamma(\nu+1/2)}\right)^p\left(\frac{16}{\phi}\right)^{p\nu}
    \re^{- \frac{p \mc{A}_{(p,r)}}{\phi}}\nn
  &\exp \left(
    p\sum_{n\geq 2}(-\phi)^{n-1}\sum_{m=0}^{[n/2]}(n+1-2m)f_{m,n+1-2m}\nu^{n-2m}
    \right),
\end{align}
and 
\begin{equation}
  \label{eq:Spr}
  S_{(p,r)} = p\sum_q S^{\text{BPS}}_{(p,q,r)}\re^{-2\pi\ri q \nu}.
\end{equation}
Since $\nu = N+1/2$ and $q\in\IZ$, the right hand side of
\eqref{eq:Spr} does not depend on the Landau level.

One important subtlety is that after using the dictionary
\eqref{eq:N-lim}, in the limit $\phi\rightarrow 0$, the flat
coordinate $t_c$ is sent to zero, and we are approaching the wall of
marginal stability across which the Borel singularities as well as the
Stokes constants change, a phenomenon related to the wall-crossing
phenomenon in 5d SYM as we mentioned at the end of
Sec.~\ref{sc:5d-SYM}.  It is then ambiguous which Stokes constants of
the perturbative Wilson loop vev should be used to compute the Stokes
constants of the perturbative Harper-Hofstadter energy series, a
question already posed in a similar contextd in \cite{Marino:2024yme}.
The answer from \cite{Marino:2024yme}, which was found emprically, is
that we should use Stokes constants from the strong coupling regime,
which corresponds to $z>1/16$ in this example.  We verify that this is
also the case here.

Let us consider the dominant Borel singularity
\begin{equation}
  \mc{A}_{(2,0)} = 16C
\end{equation}
for the perturbative energy series on the positive real axis. We recognise that
$\mc{E}^{(2,0)}(\nu,\phi) = E^{(2,0)}(\nu,\phi)$, and in particular,
this implies that
\begin{equation}
  E^{(1,0)}(\nu,\phi) = \frac{\pd E^{(0)}(\nu,\phi)}{\pd\nu}
  \frac{\sqrt{2\pi}}{\Gamma(\nu+1/2)}\left(\frac{16}{\phi}\right)^{\nu}
    \re^{- 8C/\phi}\left(1+\ldots\right),
\end{equation}
which agrees with \eqref{eq:E1} up to a trans-series coefficient, i.e.
\begin{equation}
  E^{(1)}_{\theta_x,\theta_y}(\nu,\phi) = w_{1,0}(\theta_{x,y})E^{(1,0)}(\nu,\phi)
\end{equation}
and the trans-series coefficient $w_{1,0}(\theta_{x,y})$ is given by
\begin{equation}
  w_{1,0}(\theta_{x,y}) = (-1)^{N+1}\frac{\cos\theta_x+\cos\theta_y}{\pi}
\end{equation}
as we will confirm in Sec.~\ref{sc:full-series}.
Then \eqref{eq:DelprE0} becomes
\begin{equation}
  \label{eq:Del20E0}
  \dDe{\CA_{(2,0)}} E^{(0)}(\nu,\phi) =
  \frac{S_{(2,0)}}{2\pi\ri}E^{(2,0)}(\nu,\phi).
\end{equation}
This Borel singularity can descend via
\eqref{eq:Spr} either from the three Borel singularities with
$\gamma=(2,-1,0), (2,-1\pm 1,0)$ and respective Stokes constants in
\eqref{eq:S-weak} of the perturbative Wilson loop in the weak coupling
regime, in which case, the predicted Stokes constant associated to
$\mc{A}_{(2,0)}$ is 
\begin{equation}
  S_{(2,0)}^{\text{[weak]}} = 16,
\end{equation}
or from the single Borel singularity with $\gamma = (2,0,0)$ and
Stokes constant in \eqref{eq:S-strong} of the Wilson loop in the
strong coupling regime, in which case, the predicted Stokes constant
associated to $\mc{A}_{(2,0)}$ is
\begin{equation}
  S_{(2,0)}^{\text{[strong]}} = 4.
\end{equation}
The actual numerical calculation of the Stokes discontinuity of
$E^{(0)}$ across the positive real axis compared with the right hand
side of \eqref{eq:Del20E0} shows the latter is the case,
i.e.
\begin{equation}
  \label{eq:S20}
  S_{(2,0)} = S_{(2,0)}^{\text{[strong]}} = 4.
\end{equation}
%
Another way to see that this has to be the case is notice that the
energy of the Harper-Hofstadter model has the property that $-4<E < 4$,
and this is translated to the modulus in the 5d SYM as
$z = 1/u^2 = 1/E^2 > 1/16$, which corresponds to the strong coupling
regime.

\subsection{Minimal trans-series}
\label{sc:min-series}


We would like to study the general resurgent structure of the
perturbative energy series, not only the dominant Borel singularity.
Since we will be interested in the Borel resummation with real and
positive $\phi$, we focus on the Borel singularities on the positive
real axis.  We conjecture the only Borel singularities of this type
are $\mc{A}_{(2,0)} = 16C$ and its multiples, and we will denote them
simply by
\begin{equation}
  \ell\mc{A},\quad \mc{A}:=\mc{A}_{(2,0)},\;\ell=1,2,\ldots.
\end{equation}
The action of the alien derivatives of these Borel singularities
$\dDe{k\mc{A}}$ on the perturbative energy series should follow from
the alien derivatives of the Wilson loop vev \eqref{eq:DelW}.  By a
similar calculation as in the previous section, or by simply comparing
the right hand side of \eqref{eq:DelW} with the definition of
$E^{(n,m)}$ given in \eqref{eq:Enm} together with the dictionary
\eqref{eq:tD-F}, we can conclude that
\begin{equation}
  \label{eq:DelkA-E0}
  \dDe{\ell\CA} E^{(0)}(\nu,\phi) =
  \frac{S_{\mc{A}}}{2\pi\ri}\frac{(-1)^{\ell-1}}{\ell}
  E^{(2\ell,0)}(\nu,\phi),
\end{equation}
where, as we discussed in the previous section,
\begin{equation}
  S_{\mc{A}}:= S_{(2,0)} = 4.
\end{equation}

It is also useful to consider the resurgent structure of the
trans-series building blocks $E^{(n,m)}(\nu,\phi)$.  Starting from
\eqref{eq:DelF}, and using the chain rule of alien derivatives as well
as that it commutes with ordinary derivatives, one finds
\begin{equation}
  \dDe{\ell A_{(2,0,0)}}\re^{-n\pd_{t_c}F_{\text{NS}}(t_c,\hbar)/\hbar} =
  \frac{S_{(2,0,0)}^{\text{BPS}}}{2\pi\ri}\frac{\hbar(-1)^{(\ell-1)}}{\ell}
  (2n \pd^2_{t_c}F_{\text{NS}}(t_c,\hbar))\re^{-(n+2\ell)\pd_{t_c}F_{\text{NS}}(t_c,\hbar)/\hbar}.
\end{equation}
Using this result, the Leibniz rule of alien derivatives, and
following the derivation as in the previous section, together with
\eqref{eq:Spr},\eqref{eq:S20}, one finds that
\begin{equation}
  \label{eq:DelkA-Enm}
  \dDe{\ell \CA}E^{(n,m)}(\nu,\phi) =
  \frac{S_{\CA}}{2\pi\ri}\frac{(-1)^{\ell-1}}{\ell} E^{(n+2\ell,m+1)}(\nu,\phi).
\end{equation}
It also implies that $E^{(n,m)}(\nu,\phi)$ has the same Borel
singularities as $E^{(0)}(\nu,\phi)$.

\renewcommand*{\arraystretch}{1.2}

\begin{table}
  \centering
  \begin{tabular}{*{5}{>{$}c<{$}}}\toprule
    m
    & 0 & 1 & 2 & 3\\\midrule
    v_{1,m} & -\frac{\ri}{2\pi} &&&\\
    v_{2,m} & \frac{\ri}{4\pi} & -\frac{1}{8\pi^2} &&\\
    v_{3,m}
    & -\frac{\ri}{6\pi}
        & \frac{1}{8\pi^2}
            & -\frac{\ri}{48\pi^3}&\\
    v_{4,m}
    & \frac{\ri}{8\pi} & -\frac{11}{96\pi^2} & -\frac{\ri}{32\pi^3} & \frac{1}{384\pi^4}\\
      \bottomrule
  \end{tabular}
  \caption{Trans-series coefficients $v_{n,m}$ in minimal
    trans-series.}
  \label{tab:vnm}
\end{table}

It was then argued in \cite{vanSpaendonck:2023znn} that we can include
all corrections to the energy perturbative series from all of its
Borel singularities on the positive real axis via the minimal
trans-series
\begin{equation}
  E_{\text{min}}^{(0)}(\nu,\phi;\sigma) =
  E^{(0)}(\nu,\phi) +
  \sum_{n'=1}^\infty\sum_{m'=0}^{n'-1}
  \sigma^{m'+1}v_{n',m'}E^{(2n',m')}(\nu,\phi), \label{eq:E0min}
\end{equation}
where the trans-series coefficients are
\begin{equation}
  v_{n,m} =
  \frac{1}{n!}B_{n,m+1}(1!s_1,2!s_2,\ldots,(n-m)!s_{n-m}),
\end{equation}
with $B_{n,m+1}$ being the incomplete Bell's polynomials and
\begin{equation}
  s_{j} = \frac{(-1)^{j-1}}{j}\frac{1}{2\pi\ri}, \quad j = 1,2,\ldots.
\end{equation}
The first few trans-series coefficients $v_{n,m}$ are given in
Tab.~\ref{tab:vnm}.  The minimal trans-series $E^{(0)}_{\text{min}}$
has the nice property that its Stokes automorphism across the positive
real axis is given by
\begin{equation}
  \label{eq:Stokes-E0}
  \mf{S}_0 E^{(0)}_{\text{min}}(\nu,\phi;\sigma) =
  E^{(0)}_{\text{min}}(\nu,\phi;\sigma + S_\CA).
\end{equation}
To see this, we first notice that
\begin{align}
  \mf{S}_0E^{(0)}(\nu,\hbar) =
  &\left(\exp\sum_{\ell=1}^\infty\dDe{\ell\CA}\right)E^{(0)}(\nu,\hbar)\nn
    =
  &E^{(0)} +
    \sum_{n'=1}^\infty\sum_{m'=0}^{n'-1}\frac{1}{n'!}B_{n',m'+1}(j!\dDe{j\CA})E^{(0)}\nn=
  &E^{(0)} +
    \sum_{n'=1}^\infty\sum_{m'=0}^{n'-1}S_{\CA}^{m'+1}\frac{1}{n'!}B_{n',m'+1}(j!s_j)E^{(2n',m')}\nn=
  &E^{(0)} +
    \sum_{n'=1}^\infty\sum_{m'=0}^{n'-1}S_{\CA}^{m'+1}v_{n',m'}E^{(2n',m')},
    \label{eq:SE0}
\end{align}
where from the first line to the second line we used the Fa\`a di Bruno
formula, and from the second line to the third line we have used the
resurgent properties \eqref{eq:DelkA-E0}, \eqref{eq:DelkA-Enm} as well
as the homogeneity property of incomplete Bell's polynomials
\begin{equation}
  \label{eq:B-scale}
  B_{n,m+1}(\alpha\beta x_1,\alpha^2\beta x_2,\ldots,
  \alpha^{n-m}\beta x_{n-m})
  = \alpha^n\beta^{m+1}
  B_{n,m+1}(x_1,x_2,\ldots,x_{n-m})
\end{equation}
Comparing \eqref{eq:SE0} with \eqref{eq:E0min}, we conclude that
\begin{equation}
  \label{eq:Stokes-E0}
  \mf{S}_0E^{(0)}(\nu,\hbar) = \mf{S}_0E^{(0)}_{\text{min}}(\nu,\hbar;0)
  = E^{(0)}_{\text{min}}(\nu,\hbar;S_\CA).
\end{equation}
Furthermore, let us define
\begin{equation}
  \dDes{\ell\CA} E^{(n,m)} = \frac{(-1)^{\ell-1}}{\ell} E^{(n+2\ell,m+1)}
\end{equation}
so that $\dDe{\ell\CA} = S_{\CA}\dDes{\ell\CA}$ and
\begin{equation}
  \mf{S}_0[S_\CA] = \exp\left(S_\CA\sum_{\ell=1}^\infty\dDes{\ell\CA}\right).
\end{equation}
Since
\begin{equation}
  \mf{S}_0[S_1]\mf{S}_0[S_2] = \mf{S}_0[S_1+S_2],
\end{equation}
we finally arrive at
\begin{align}
  \mf{S}_0E^{(0)}_{\text{min}}(\nu,\phi;\sigma) =
  &\mf{S}_0[S_\CA]\mf{S}_0[\sigma]E^{(0)}_{\text{min}}(\nu,\phi;0)  =
    \mf{S}_0[S_\CA+\sigma]E^{(0)}_{\text{min}}(\nu,\phi;0) \nn=
  &E^{(0)}_{\text{min}}(\nu,\phi;\sigma+S_{\CA}).
    \label{eq:Stokes-Enmsigma}
\end{align}

\begin{table}
  \centering
  \begin{tabular}{*{3}{>{$}c<{$}}}\toprule
    n
    & \mr{S}^{(+)}E_{\text{min}}^{(0)}(0,\phi,-2)
    &\mr{S}^{(-)}E_{\text{min}}^{(0)}(0,\phi,+2)\\
    \midrule
    0 & 3.545 + 3.794\times 10^{-12}\ri & 3.545 - 3.794\times 10^{-12}\ri\\
    2 & 3.545 - 2.485\times 10^{-23}\ri & 3.545 + 2.485\times 10^{-23}\ri\\
    4 & 3.545 - 6.074\times 10^{-33}\ri & 3.545 + 6.074\times 10^{-33}\ri\\
    6 & 3.545 - 2.074\times 10^{-38}\ri & 3.545 + 2.074\times 10^{-38}\ri\\
    \bottomrule
  \end{tabular}
  \caption{Borel resummation of minimal trans-series
    $E_{\text{min}}^{(0)}(N,\phi;\mp 2)$ at Landau level $N=0$ with
    $\phi = 2\pi/13$.  $n$ is the level of instanton corrections
    included.  As higher level instanton corrections are included, the
    imaginary part of the resummation becomes smaller.}
  \label{tab:min-13-N0}
\end{table}

\begin{table}
  \centering
  \begin{tabular}{*{3}{>{$}c<{$}}}\toprule
    n
    & \mr{S}^{(+)}E_{\text{min}}^{(0)}(0,\phi,-2)
    &\mr{S}^{(-)}E_{\text{min}}^{(0)}(0,\phi,+2)\\
    \midrule
    0 & 3.736 + 2.985\times 10^{-22}\ri & 3.736 - 2.985\times 10^{-22}\ri\\
    2 & 3.736 - 2.651\times 10^{-43}\ri & 3.736 + 2.651\times 10^{-43}\ri\\
    4 & 3.736 + 2.247\times 10^{-60}\ri & 3.736 - 2.247\times 10^{-60}\ri\\
    \bottomrule
  \end{tabular}
  \caption{Borel resummation of minimal trans-series
    $E_{\text{min}}^{(0)}(N,\phi;\mp 2)$ at Landau level $N=0$ with
    $\phi = 2\pi/23$.  $n$ is the level of instanton corrections
    included.  As higher level instanton corrections are included, the
    imaginary part of the resummation becomes smaller.}
  \label{tab:min-23-N0}
\end{table}

\begin{table}
  \centering
  \begin{tabular}{*{3}{>{$}c<{$}}}\toprule
    n
    & \mr{S}^{(+)}E_{\text{min}}^{(0)}(1,\phi,-2)
    &\mr{S}^{(-)}E_{\text{min}}^{(0)}(1,\phi,+2)\\
    \midrule
    0 & 2.691 + 3.190\times 10^{-9}\ri  & 2.691 - 3.190\times 10^{-9}\ri\\
    2 & 2.691 - 2.001\times 10^{-17}\ri & 2.691 + 2.001\times 10^{-17}\ri\\
    4 & 2.691 - 2.454\times 10^{-24}\ri & 2.691 + 2.454\times 10^{-24}\ri\\
    6 & 2.691 + 7.118\times 10^{-32}\ri & 2.691 - 7.118\times 10^{-32}\ri\\
    \bottomrule 
  \end{tabular}
  \caption{Borel resummation of minimal trans-series
    $E_{\text{min}}^{(0)}(N,\phi;\mp 2)$ at Landau level $N=1$ with
    $\phi = 2\pi/13$.  $n$ is the level of instanton corrections
    included.  As higher level instanton corrections are included, the
    imaginary part of the resummation becomes smaller.}
  \label{tab:min-13-N1}
\end{table}

\begin{table}
  \centering
  \begin{tabular}{*{3}{>{$}c<{$}}}\toprule
    n
    & \mr{S}^{(+)}E_{\text{min}}^{(0)}(1,\phi,-2)
    &\mr{S}^{(-)}E_{\text{min}}^{(0)}(1,\phi,+2)\\
    \midrule
    0 & 3.226 + 8.873\times 10^{-19}\ri & 3.226 - 8.873\times 10^{-19}\ri\\
    2 & 3.226 - 2.514\times 10^{-36}\ri & 3.226 + 2.514\times 10^{-36}\ri\\
    4 & 3.226 - 2.138\times 10^{-52}\ri & 3.226 + 2.138\times 10^{-52}\ri\\
    \bottomrule
  \end{tabular}
  \caption{Borel resummation of minimal trans-series
    $E_{\text{min}}^{(0)}(N,\phi;\mp 2)$ at Landau level $N=1$ with
    $\phi = 2\pi/23$.  $n$ is the level of instanton corrections
    included.  As higher level instanton corrections are included, the
    imaginary part of the resummation becomes smaller.}
  \label{tab:min-23-N1}
\end{table}

In the example of the Harper-Hofstadter model, with $S_\CA =4$ for
the Borel singularities $\CA = 16C$, this implies that there is an
ambiguity-free prescription of performing Borel resummation of the
minimal trans-series 
\begin{equation}
  \label{eq:Emin-sum}
  \mr{S}^{(+)}E_{\text{min}}^{(0)}(\nu,\phi;-2)  =
  \mr{S}^{(-)}E_{\text{min}}^{(0)}(\nu,\phi;+2),
\end{equation}
%
which has the additional nice property that it is a real value, in
contrast to lateral resummations of $E^{(0)}(\nu,\phi)$ which are
always complex.  Some numerical evidences are provided in
Tabs.~\ref{tab:min-13-N0}, \ref{tab:min-23-N0}, \ref{tab:min-13-N1},
\ref{tab:min-23-N1}.

For later purpose, we will also introduce the minimal trans-series for
the building blocks $E^{(n,m)}$ and they read
\begin{align}
  E_{\text{min}}^{(n,m)}(\nu,\phi;\sigma) :=
  E^{(n,m)}(\nu,\phi) +
  \sum_{n'=1}^\infty\sum_{m'=0}^{n'-1}
  \sigma^{m'+1}v_{n',m'}E^{(n+2n',m+m'+1)}(\nu,\phi).
\end{align}
Using a similar argument with \eqref{eq:DelkA-Enm}, one can show that
across the positive real axis
\begin{equation}
  \label{eq:Stokes-Enm}
  \mf{S}_0 E_{\text{min}}^{(n,m)}(\nu,\phi;\sigma) =
  E_{\text{min}}^{(n,m)}(\nu,\phi;\sigma+S_\CA).
\end{equation}

\subsection{Full trans-series and the exact quantization condition}
\label{sc:full-series}


The minimal trans-series $E_{\text{min}}^{(0)}(\nu,\phi;\sigma)$
encodes the minimal resurgent structure starting from
$E^{(0)}(\nu,\phi)$ accessible via Borel singularities on the positive
real axis.  If the strong resurgence program were to hold here, it
would be the entire story, and the Borel resummation
\eqref{eq:Emin-sum} would be the exact energy spectrum.  But as we
have discussed in Sec.~\ref{sc:summability}, it misses at least the
1-instanton sector, and the full energy trans-series would be a
superset of the minimal energy trans-series.

The way to construct a larger and full trans-series which includes the
minimal trans-series as a consistent component is via the procedure of
``tensor product'' of trans-series introduced in
\cite{vanSpaendonck:2023znn}.  We assume that the full energy
trans-series still has the form of \eqref{eq:E-trans-gen}.  Suppose
the right hand side of \eqref{eq:DeltR} can split into the sum of two
functions
\begin{equation}
  R(\lambda) = R^A(\lambda) + R^B(\lambda) =
  \sum_{k=1}^\infty (r_k^A+r_k^B)\lambda^k
\end{equation}
where we have replaced $\hbar$ by $\phi$, we can define two implicit
equations\footnote{Note that $\Delta\nu$ as solution to
  \eqref{eq:DeltR} is not equal to the sum of
  $\Delta\nu_A$,$\Delta\nu_B$ as solutions to
  \eqref{eq:DeltRA},\eqref{eq:DeltRB}.}
\begin{align}
  &\Delta\nu_A = R^A(\lambda(\nu+\Delta\nu_A)),\label{eq:DeltRA}\\
  &\Delta\nu_B = R^B(\lambda(\nu+\Delta\nu_B)),\label{eq:DeltRB}
\end{align}
with $\lambda(\nu) = \re^{-\frac{1}{2\phi}t_D(\nu,\phi)}$, and the
solutions define respectively two sets of trans-series:
the trans-series of $A$ type
\begin{subequations}
\begin{align}
  E^{(0)}_A(\nu,\phi) =
  &\re^{-\nu_A\pd_\nu}E^{(0)}(\nu,\phi) \nn=
  &E^{(0)}(\nu,\phi)  +
    \sum_{n'=1}^\infty\sum_{m'=0}^{n'-1} u_{n',m'}(\underline{r}^A) E^{(n',m')}(\nu,\phi),\\
  E^{(n,m)}_A(\nu,\phi) =
  &\re^{-\nu_A\pd_\nu}E^{(n,m)}(\nu,\phi) \nn=
  &E^{(n,m)}(\nu,\phi)  +
    \sum_{n'=1}^\infty\sum_{m'=0}^{n'-1} u_{n',m'}(\underline{r}^A) E^{(n+n',m+m'+1)}(\nu,\phi),
\end{align}
\end{subequations}
and the trans-series of $B$ type
\begin{subequations}
\begin{align}
  E^{(0)}_B(\nu,\phi) =
  &\re^{-\nu_A\pd_\nu}E^{(0)}(\nu,\phi) \nn=
  &E^{(0)}(\nu,\phi)  +
    \sum_{n'=1}^\infty\sum_{m'=0}^{n'-1} u_{n',m'}(\underline{r}^B) E^{(n',m')}(\nu,\phi),\\
  E^{(n,m)}_B(\nu,\phi) =
  &\re^{-\nu_A\pd_\nu}E^{(n,m)}(\nu,\phi) \nn=
  &E^{(n,m)}(\nu,\phi)  +
    \sum_{n'=1}^\infty\sum_{m'=0}^{n'-1} u_{n',m'}(\underline{r}^B) E^{(n+n',m+m'+1)}(\nu,\phi),
\end{align}
\end{subequations}
and the full trans-series
\begin{align}
  \label{eq:full-1}
  E(\nu,\phi) = 
  E^{(0)}(\nu,\phi) +
  \sum_{n=1}^\infty\sum_{m=0}^{n-1}u_{n,m}(\underline{r})E^{(n,m)}(\nu,\phi),
\end{align}
can be formulated as the ``tensor product'',
\begin{equation}
  \text{full trans-series} \simeq \text{trans-series}\;A \otimes
  \text{trans-series}\;B,
\end{equation}
in the sense that it can be equally written as
\begin{equation}
  \label{eq:full-2}
  E(\nu,\phi) = E^{(0)}_A(\nu,\phi) + \sum_{n=1}^\infty\sum_{m=0}^{n-1}u_{n,m}(\underline{r}^B)E^{(n,m)}_A(\nu,\phi).
\end{equation}
This can be understood as arising from a two step application of
\eqref{eq:E-E0Del}, \eqref{eq:E-trans-gen},
\begin{align} 
  E(\nu,\phi) =
  &e^{\Delta\nu_A\partial\nu}e^{\Delta\nu_B\partial\nu}E^{(0)}(\nu) =
    E^{(0)}(\nu+\Delta\nu^A)
    +\sum_{n=1}^\infty\sum_{m=0}^{n-1}u_{n,m}(\und{r}^B)E^{(n,m)}(\nu+\Delta\nu^A,\phi)\nn
    =&E^{(0)}_{A}(\nu)
       +\sum_{n=1}^\infty\sum_{m=0}^{n-1}u_{n,m}(\und{r}^B)E^{(n,m)}_A(\nu,\phi),
\end{align}
and it can be verified by checking the identities of trans-series
coefficients
\begin{align}
  \label{eq:unmtensorfac}
  u_{n,m}(\und{r}) =
  &u_{n,m}(\und{r}^A) + u_{n,m}(\und{r}^B)
    +
    \sum_{n'=1}^{n-1}\sum_{m'=\text{max}(m-n+n',0)}^{\text{min}(m-1,n'-1)}
    u_{n-n',m-m'-1}(\und{r}^B)u_{n',m'}(\und{r}^A),\nn
    \quad
  &n = 1,2,\ldots,\infty,\; m=0,1,\ldots,n-1,
\end{align}
obtained by comparing coefficients of \eqref{eq:full-1} and
\eqref{eq:full-2}.

In the case of the Harper-Hofstadter model, we first notice that the
minimal trans-series \eqref{eq:E0min} can be put in the form of
\eqref{eq:E-trans-gen} with
\begin{equation}
  u_{n,m}(\underline{r}^{\text{min}}) =
  \begin{cases}
    v_{n/2,m},\quad  &\text{even}\;n,\\
    0,\quad &\text{odd}\;n,
  \end{cases}\quad
  r_j^{\text{min}} =
  \begin{cases}
    s_{j/2},\quad &\text{even}\; j,\\
    0,\quad &\text{odd}\; j.
  \end{cases}
\end{equation}
Therefore, the minimal trans-series can be written as
\begin{equation}
  E_{\text{min}}^{(0)}(\nu,\phi;\sigma) = E^{(0)}(\nu+ \sigma \Delta \nu^{\text{min}},\phi)
\end{equation}
where $\Delta \nu^{\text{min}}$ is solution to
\begin{equation}
  \Delta \nu^{\text{min}}=  \sum_{j=1}^\infty s_j \lambda^{2j}  =:
  R^{\text{min}}(\lambda),\quad \lambda =
  \re^{-\frac{1}{2\phi}t_D(\nu+\Delta \nu^{\text{min}},\phi)}
\end{equation}

Now without loss of generality, we can assume that for the
Harper-Hofstadter model, the right hand side of \eqref{eq:DeltR} can
indeed be split as
\begin{equation}
  \Delta \nu = \sigma R^{\text{min}}(\lambda) +
  R^{\text{med}}(\lambda)
  = \sum_{j\geq 1} (\sigma r_j^{\text{min}} + r_j^{\text{med}})\lambda^j
\end{equation}
where $r_j^{\text{med}}$ are yet unknown.
Then the full trans-series can be written as
\begin{align}
  E_{\theta_{x,y},\sigma}(\nu,\phi) =
  E_{\text{min}}^{(0)}(\nu,\phi;\sigma) + \sum_{n=1}^\infty
  \sum_{m=0}^{n-1}w_{n,m}(\theta_{x,y})
  E_{\text{min}}^{(n,m)}(\nu,\phi;\sigma),\label{eq:tensor}
\end{align}
where we have denoted the trans-series coefficients
\begin{equation}
  w_{n,m}:= u_{n,m}(\underline{r}^{\text{med}}(\theta_{x,y}))
\end{equation}
which depend on the Bloch angles $\theta_x,\theta_y$.

\begin{figure}
  \centering
  \subfloat[$\phi=2\pi/13$]{\includegraphics[height=4cm]{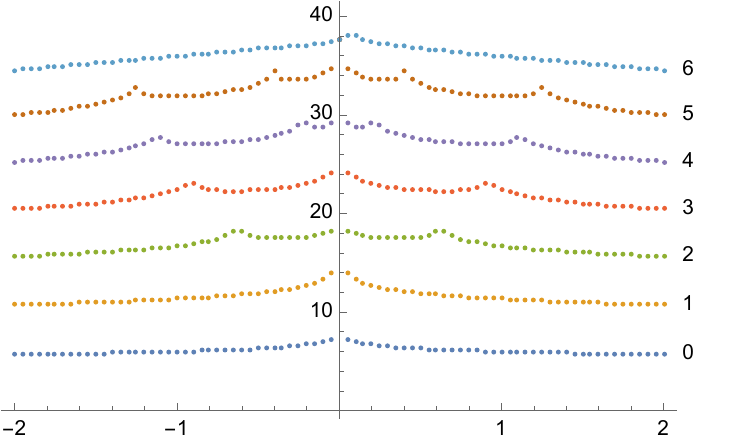}}\hspace{2ex}
  \subfloat[$\phi=2\pi/23$]{\includegraphics[height=4cm]{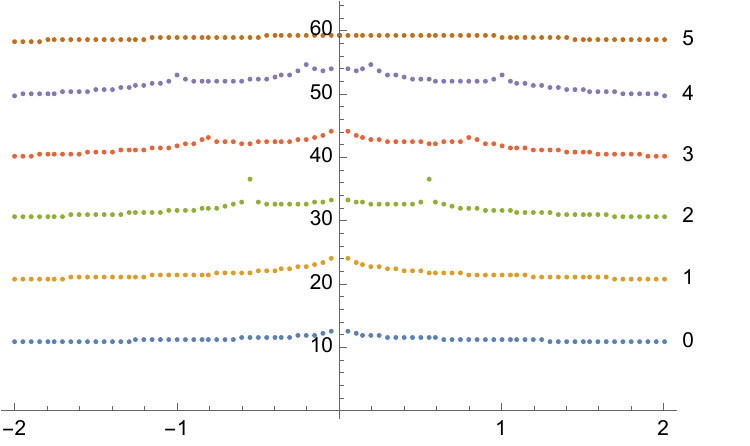}}
  \caption{The order of magnitude ($-\log_{10}(|*|)$, vertical axis)
    of the difference between the exact spectrum and the Borel
    resummation of full energy trans-series in the form of
    \eqref{eq:tensor} at Landau level 0 with varying $\Theta$
    (horizontal axis).  We include progressively contributions of
    increasing instanton orders $n=0,1,2,\ldots$ from lower data
    points to higher data points.}
  \label{fig:spec-diff-P1lev0}
\end{figure}

\begin{figure}
  \centering
  \subfloat[$\phi=2\pi/13$]{\includegraphics[height=4cm]{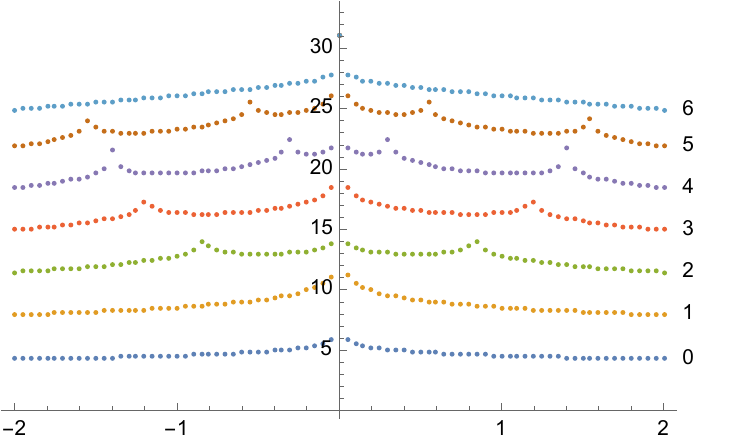}}\hspace{2ex}
  \subfloat[$\phi=2\pi/23$]{\includegraphics[height=4cm]{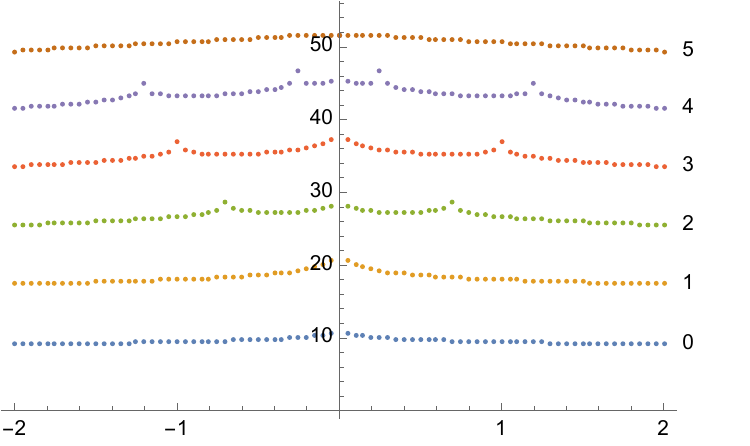}}
  \caption{The order of magnitude ($-\log_{10}(|*|)$, vertical axis)
    of the difference between the exact spectrum and the Borel
    resummation of full energy trans-series in the form of
    \eqref{eq:tensor} at Landau level 1 with varying $\Theta$
    (horizontal axis).  We include progressively contributions of
    increasing instanton orders $n=0,1,2,\ldots$ from lower data
    points to higher data points.}
  \label{fig:spec-diff-P1lev1}
\end{figure}

The weak resurgence program dictates that in the regime $\phi\nu\ll1$,
the exact energy spectrum is given by 
\begin{equation}
  E^{\text{ext}}_{\theta_{x},\theta_y}(\nu,\phi) =
  \mr{S}^{(+)}E_{\theta_{x},\theta_y,-2}(\nu,\phi)
  =
  \mr{S}^{(-)} E_{\theta_{x},\theta_y,+2}(\nu,\phi).
\end{equation}
%
The resurgent properties \eqref{eq:Stokes-E0} and
\eqref{eq:Stokes-Enm} make sure that the two prescriptions of lateral
Borel resummation yield the same result.  This gives us a method to
fix the unknown trans-series coefficients $w_{n,m}$.  By comparing
with the exact energy spectrum solved from the secular equation
\eqref{eq:secular1} at $\phi = 2\pi/Q$, with high precision numerical
calculations, we find the first few trans-series coefficients $w_{n,m}$
up to $n=6$, i.e. up to 6-instanton order, as tabulated in
Tab.~\ref{tab:wnm}.
Here we have introduced notation
\begin{equation}
  \label{eq:Theta}
  \Theta:= (-1)^{N+1}(\cos\theta_x+\cos\theta_y).
\end{equation}
Some numerical evidences are provided in
Fig.~\ref{fig:spec-diff-P1lev0}, \ref{fig:spec-diff-P1lev1}.  It turns
out that these trans-series coefficients can indeed be written in the
form of \eqref{eq:uB}.  In fact, we find the trans-series coefficients
$w_{n,m}$ can be expressed as
\begin{equation}
  \label{eq:wnm}
  w_{n,m} = \frac{1}{n!}B_{n,m+1}(1!t_1,2!t_2,\ldots,(n-m)!t_{n-m})
\end{equation}
where the parameters $t_j$ are such that the generating function for
$t_j$ is
\begin{equation}
  \label{eq:tj}
  \sum_{j\geq 1}t_j\lambda^j =
  \frac{1}{\pi}\arcsin\frac{\Theta}{\lambda+\lambda^{-1}}.
\end{equation}
%
This in turn validates our conjecture that the full energy
trans-series can be written as a tensor product of the minimal
trans-series and a secondary trans-series, which is called the medium
in the sense of \eqref{eq:full-2}.

Note that the medium trans-series coefficients all have the property that
they vanish in the van Hove singularity with $\Theta = 0$.  In particular this
implies that 
\begin{equation}
  E^{\text{ext}}_{0,0}(\nu,\phi)  =
  \mr{S}^{(+)}E_{\text{min}}^{(0)}(\nu,\phi;-2) =
  \mr{S}^{(-)}E_{\text{min}}^{(0)}(\nu,\phi;+2).
\end{equation}
Moreover, taking the difference between full trans-series evaluated at
$\Theta=2$ and $\Theta=-2$, one finds the exact formula for the energy
bandwidths at $P=1$ to be 
\begin{equation}
  \mathrm{bw}^{\text{ext}}_N(\phi) =
  2\sum_{n=1}^{\infty}\sum_{m=0}^{n-1}w_{2n-1,m}(\Theta=2)
  \mr{S}^{(\pm)}E_{\text{min}}^{(2n-1,m)}(N+\frac{1}{2},\phi;\mp 2).
\end{equation}
      
Once we have calculated the coefficients $w_{n,m}$ for the medium
trans-series, we can use \eqref{eq:unmtensorfac} to build the
coefficients $u_{n,m}$ for the full trans-series from $v_{n,m}$ and
$w_{n,m}$.  The first few examples are in Tab.~\ref{tab:unm}.  They
reduce to $v_{n,m}$ if we set $\Theta = 0$ and reduce to $w_{n,m}$ if
we set $\ep = 0$.  Alternatively, we can add up the generating series
for $s_j$ and $t_j$ and find the generating series for the parameter
$r_j$ of the full trans-series,
\begin{equation}
  \sum_{j\geq 1} r_j\lambda^j =
  \frac{\ri}{\epsilon\pi}\log\left(\sqrt{1+(2-\Theta^2)\lambda^2+\lambda^4}
    -\ri\epsilon\Theta\lambda\right).
\end{equation}
Taking the logic in Sec.~\ref{sc:WKB} backwards, this implies the EQCs
\begin{equation}
  \label{eq:EQC-HH}
  D_{\theta_{x},\theta_y}^{\pm}:\quad
  1+\CV_A^{\pm 1}(1+\CV_B)^2 - 2\sqrt{\CV_A^{\pm 1}
    \CV_B}\,\Theta = 0.
\end{equation}
The two conditions $D_{\theta_{x},\theta_y}^{\pm}$ are suitable for
the two choices of the lateral Borel resummations $\mr{S}^{(\pm)}$
respectively.  These two quantization conditions lead to the same
energy spectrum as they are correctly related by the Stokes
transformation of the Voros symbols.  As explained in
\cite{Gu:2022fss,Gu:2023wum}, the Stokes transforms of Voros symbols
are controled by the BPS invariants of the corresponding
supersymmetric field theory
\begin{equation}
  \mf{S}_\theta: \quad \mc{V}_A \rightarrow
  \mc{V}_A(1+\mc{V}_B)^{\vev{\gamma_A,\gamma_B}\Omega(\gamma_B)},
\end{equation}
where we take the convention for the Dirac pairing of EM charges,
\begin{equation}
  \vev{\gamma_A,\gamma_B} = p_B q_A - p_A q_B,\quad
  \gamma_A=(p_A,q_A,r_A),\; \gamma_B=(p_B,q_B,r_B).
\end{equation}
As we discussed in the previous sections, the supersymmetric field
theory corresponding to the Harper-Hofstader model is the 5d SYM on
$S^1\times\IR^4$ in the strong coupling regime.  The charge vectors
associated to $\mc{V}_A, \mc{V}_B$ are respectively
\begin{equation}
  \gamma_A = (0,1,0), \quad\gamma_B = (2,0,0),
\end{equation}
with
\begin{equation}
  \vev{\gamma_A,\gamma_B} = 2,\quad \Omega(\gamma_B) = 2
\end{equation}
as discussed in Sec.~\ref{sc:5d-SYM}, so that
\begin{equation}
  \mf{S}_0: \quad \mc{V}_A \rightarrow
  \mc{V}_A(1+\mc{V}_B)^{4},
\end{equation}
which makes sure that the two conditions in \eqref{eq:EQC-HH} are
equivalent to each other.  Note that this is different from Mathieu
equation, where the corresponding supersymmetric field theory is 4d
SYM, and the BPS invariant $\Omega(\gamma_B) = 1$.  This implies a
slightly different form of Stokes transformation of Voros symbols
\begin{equation}
    \mf{S}_0: \quad \mc{V}_A \rightarrow
  \mc{V}_A(1+\mc{V}_B)^{2},
\end{equation}
which also makes sure that the two forms of EQC in
\eqref{eq:EQC-cosine} are equivalent to each other.



If we introduce the medium resummation
\begin{equation}
  \mr{S}^{(\text{med})} = \frac{1}{2}\left(\mr{S}^{(+)}+\mr{S}^{(-)}\right),
\end{equation}
the EQC can be written as 
\begin{equation}
  D_{\theta_{x},\theta_{y}}^{\text{med}}:\quad (1+\CV_A)(1+\CV_B)
  -2\sqrt{\CV_A\CV_B}\,\Theta  = 0.
\end{equation}
which is more symmetric between the perturbative and the
non-perturbative Voros symbols.  Note that the medium trans-series
with coefficients \eqref{eq:wnm} can be solved from this quantization
condition, which explains its name.

\begin{table}
  \centering
  \begin{tabular}{*{7}{>{$}c<{$}}}\toprule
    m
    & 0 & 1 & 2 & 3 & 4 & 5\\\midrule
    w_{1,m} & \frac{\Theta}{\pi} &&&&&\\
    w_{2,m} & 0 & \frac{\Theta^2}{2\pi^2} &&&&\\
    w_{3,m}
    & -\frac{\Theta}{\pi} + \frac{\Theta^3}{6\pi}
        & 0
            & \frac{\Theta^3}{6\pi^3}&&&\\
    w_{4,m}
    & 0 & -\frac{\Theta^2}{\pi^2} + \frac{\Theta^4}{6\pi^2}
            & 0 & \frac{\Theta^4}{24\pi^4}&&\\
    w_{5,m}
    & \frac{\Theta}{\pi}-\frac{\Theta^3}{2\pi} + \frac{3\Theta^5}{40\pi}
        & 0 & -\frac{\Theta^3}{2\pi^3}+\frac{\Theta^5}{12\pi^3} & 0
                    & \frac{\Theta^5}{120\pi^5}&\\
    w_{6,m}
    & 0 & \frac{3\Theta^2}{2\pi^2}-\frac{2\Theta^4}{3\pi^2} + \frac{4\Theta^6}{45\pi^2}
        & 0 & -\frac{\Theta^4}{6\pi^4}+\frac{\Theta^6}{36\pi^4} & 0
                    & \frac{\Theta^6}{720\pi^6}\\
    \bottomrule
  \end{tabular}
  \caption{Trans-series coefficients $w_{n,m}$ in medium
    trans-series.}
  \label{tab:wnm}
\end{table}

\begin{table}
  \centering
  \resizebox{\linewidth}{!}{
  \begin{tabular}{*{7}{>{$}c<{$}}}\toprule
    m
    & 0 & 1 & 2 & 3 & 4 & 5\\\midrule
    u_{1,m} & \frac{\Theta}{\pi} &&&&&\\
    u_{2,m} & \frac{\ri\epsilon}{\pi} & \frac{\Theta^2}{2\pi^2} &&&&\\
    u_{3,m}
    & -\frac{\Theta}{\pi} + \frac{\Theta^3}{6\pi}
        & \frac{\ri\epsilon\Theta}{\pi^2}
            & \frac{\Theta^3}{6\pi^3}&&&\\
    u_{4,m}
    & -\frac{\ri\epsilon}{2\pi} & -\frac{\epsilon^2}{2\pi^2}-\frac{\Theta^2}{\pi^2} + \frac{\Theta^4}{6\pi^2}
            & \frac{\ri\epsilon\Theta}{2\pi^3} & \frac{\Theta^4}{24\pi^4}&&\\
    u_{5,m}
    & \frac{\Theta}{\pi}-\frac{\Theta^3}{2\pi} + \frac{3\Theta^5}{40\pi}
        & -\frac{3\ri\epsilon\Theta}{2\pi^2}+ \frac{\ri\epsilon\Theta^3}{6\pi^2}
            & -\frac{\Theta^2}{2\pi^3}-\frac{\Theta^3}{2\pi^3}+\frac{\Theta^5}{12\pi^3}
                & \frac{\ri\epsilon\Theta^3}{6\pi^4}
                    & \frac{\Theta^5}{120\pi^5}&\\
    u_{6,m}
    & \frac{\ri\epsilon}{3\pi}
        & \frac{1}{2\pi^2}+\frac{3\Theta^2}{2\pi^2}-\frac{2\Theta^4}{3\pi^2} + \frac{4\Theta^6}{45\pi^2}
            & -\frac{\ri\epsilon}{6\pi^3}-\frac{5\ri\epsilon\Theta^2}{4\pi^3}+\frac{\ri\epsilon\Theta^4}{6\pi^3}
                & -\frac{\Theta^2}{4\pi^4}-\frac{\Theta^4}{6\pi^4}+\frac{\Theta^6}{36\pi^4}
                    & \frac{\ri\epsilon\Theta^4}{24\pi^5}
                        & \frac{\Theta^6}{720\pi^6}\\
    \bottomrule
  \end{tabular}}
  \caption{Trans-series coefficients $u_{n,m}$ for the full energy
    trans-series.}
  \label{tab:unm}
\end{table}

\section{Characterization of splitting bands}
\label{sc:Csb}

\begin{figure}
  \centering%
  \includegraphics[height=7cm,valign=c]{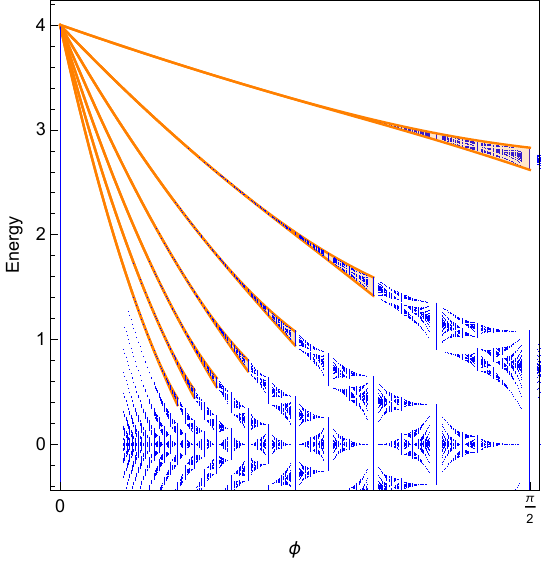}\hspace{2ex}%
  \includegraphics[height=7.4cm,valign=c]{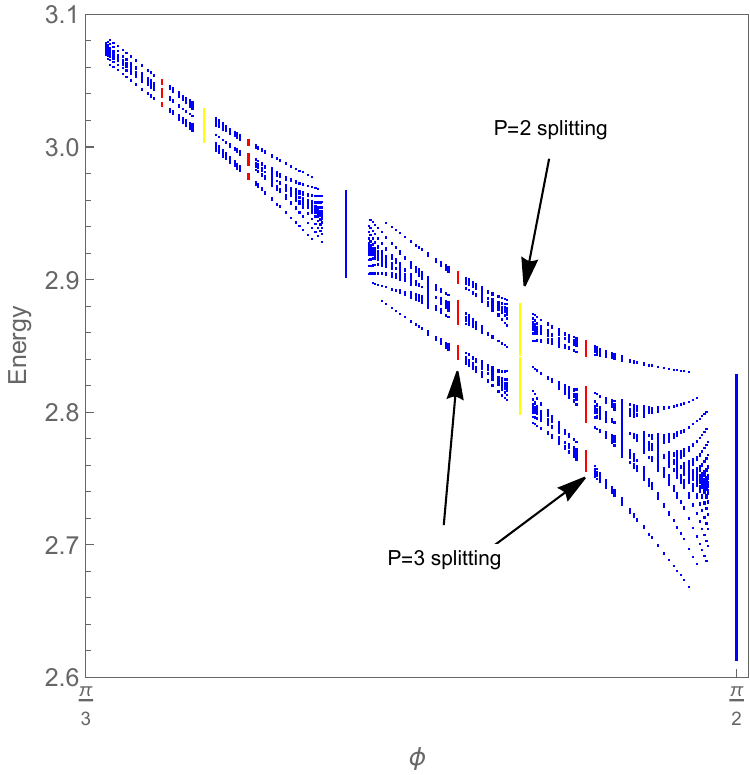}%
  \caption{Left: Hofstadter's butterfly with energy trans-series for
    $P=1$ up to the sixth Landau level. Right: zooming in on the
    lowest Landau level. We depict $P=2$ band splitting in yellow and
    $P=3$ band splitting in red.} 
  \label{fig:splitting bands}
\end{figure}

The result of the last section provides an alternative quantization
method for the Harper-Hofstadter model with flux $\phi=2\pi/Q$. From
the left graph of Fig.~\ref{fig:splitting bands}, we can easily tell
that this approach is valid pretty well into the non-perturbative
regime. In fact, we have checked that the alternative quantization
method is valid for $Q\geq 2N+3$.  For $P>1$, an important difference
from the $P=1$ case is that a single energy band at $\phi = 2\pi P/Q$,
which we call the primary Landau level, splits to $P$ smaller
secondary energy bands, which is also visible on the right graph of
Fig.~\ref{fig:splitting bands}. How to characterize this phenomenon
would be the main goal of this section.

\subsection{Self-similarity of the butterfly revisited }
\label{sc:P>1}


The discussion of the resurgent properties of the energy trans-series
as well as the construction of minimal trans-series in
Sec.~\ref{sc:min-series} is universal and it holds true for any
rational value of $\phi$.  The construction of the medium trans-series
and the consequent matching with the exact energy spectrum, however,
depends surprisingly on the numerator of $\phi = 2\pi P/Q$.  If the
flux $\phi$ is such that $P>1$, the generating series \eqref{eq:wnm}
with \eqref{eq:tj} are no longer valid.  In fact, the coefficients
$w_{n,m}$ for the medium trans-series become vastly more complicated.

\begin{figure}[t]
  \centering
  \subfloat[$\phi/(2\pi)=2/49$]{\includegraphics[height=4cm]{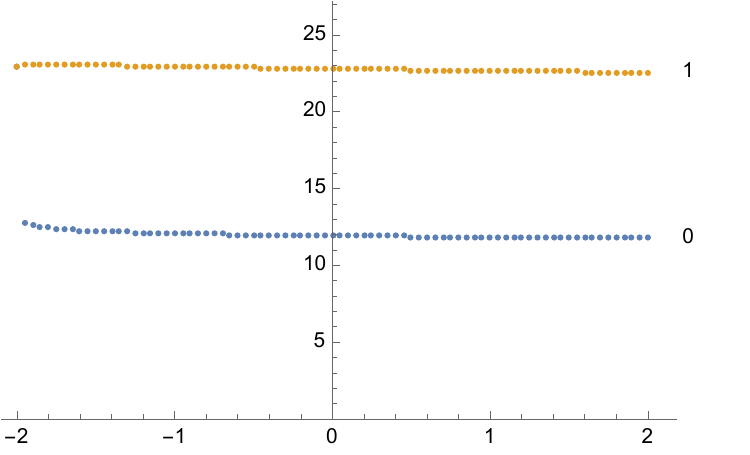}}%
  \hspace{2ex}%
  \subfloat[$\phi/(2\pi)=3/73$]{\includegraphics[height=4cm]{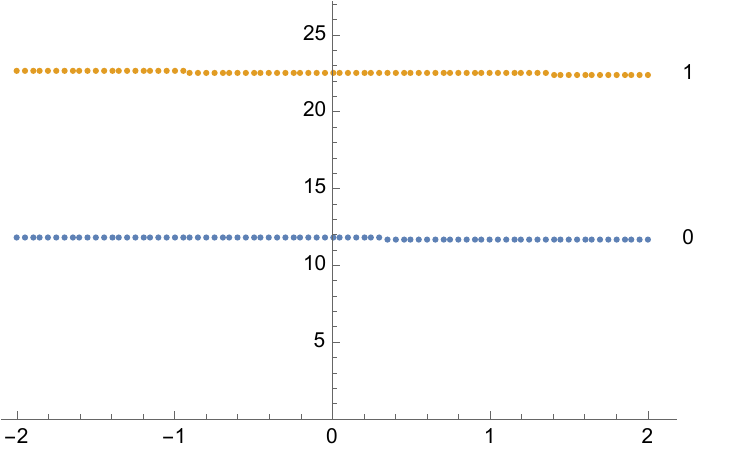}}%
  \hspace{2ex}%
  \subfloat[$\phi/(2\pi)=4/97$]{\includegraphics[height=4cm]{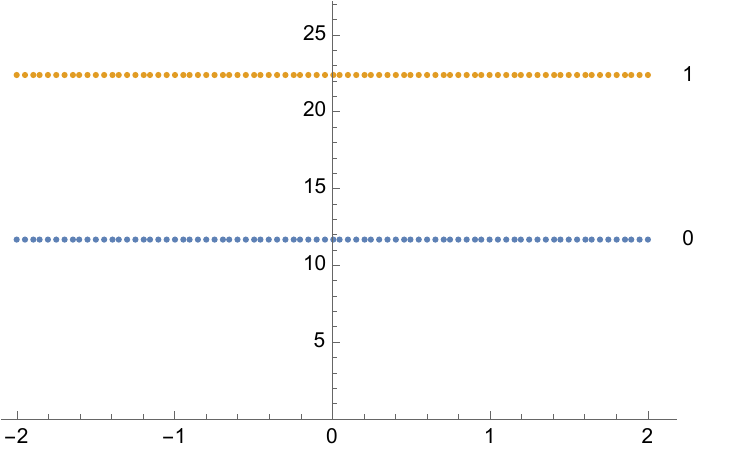}}%
  \hspace{2ex}%
  \subfloat[$\phi/(2\pi)=5/121$]{\includegraphics[height=4cm]{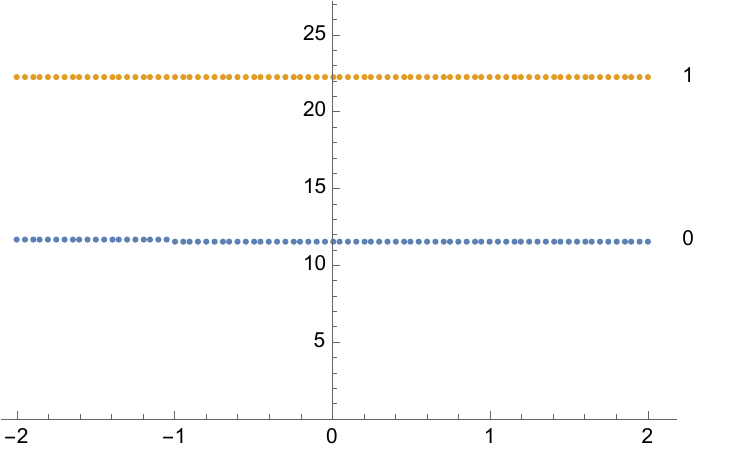}}%
  \hspace{2ex}%
  \subfloat[$\phi/(2\pi)=5/122$]{\includegraphics[height=4cm]{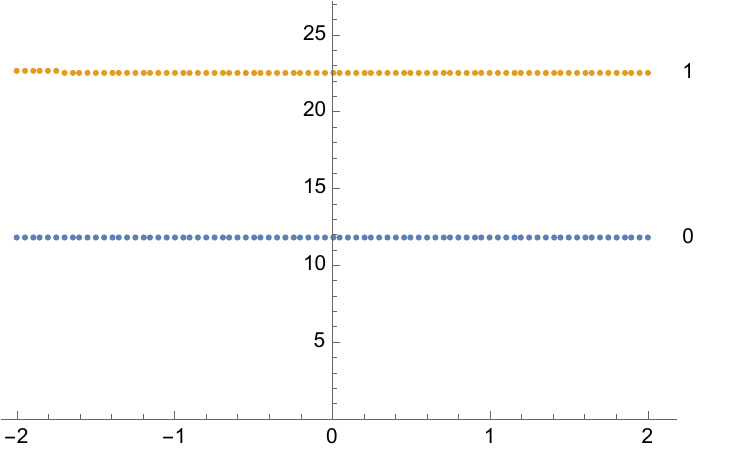}}%
  \hspace{2ex}%
  \subfloat[$\phi/(2\pi)=6/145$]{\includegraphics[height=4cm]{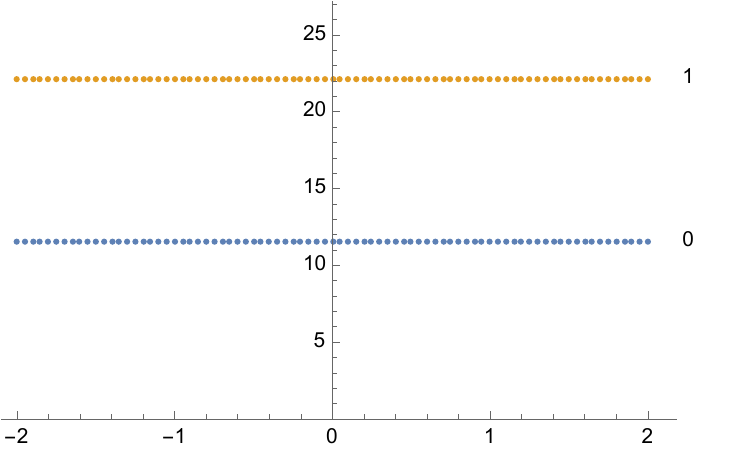}}%
  \caption{The orders of magnitude ($-\log_{10}(|*|)$, vertical axis)
    of the difference between the exact spectrum and the Borel
    resummation of full energy trans-seriesin the form of
    \eqref{eq:tensor} at Landau level 0 with varying $\Theta$
    (horizontal axis).  We include progressively contributions of
    increasing instanton orders $n=0,1$ from lower data points to
    higher data points.  The six plots are examples of (a)
    $P=2,Q=1\;\text{mod}\;P$ (a) $P=3,Q=1\;\text{mod}\;P$ (a)
    $P=4,Q=1\;\text{mod}\;P$ (a) $P=5,Q=1\;\text{mod}\;P$ (a)
    $P=5,Q=2\;\text{mod}\;P$ (a) $P=6,Q=1\;\text{mod}\;P$ for
    $\phi = 2\pi P/Q$.}
  \label{fig:spec-diff-PQ-lev0}
\end{figure}

\begin{figure}[t]
	\centering
	\subfloat[$\phi/(2\pi)=2/49$]{\includegraphics[height=4cm]{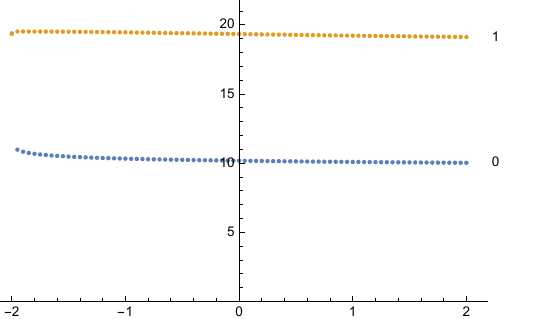}}%
	\hspace{2ex}%
	\subfloat[$\phi/(2\pi)=3/73$]{\includegraphics[height=4cm]{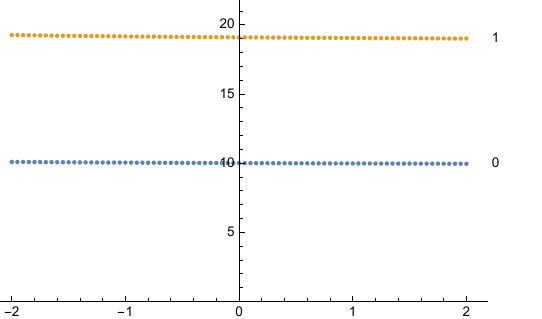}}%
	\hspace{2ex}%
	\subfloat[$\phi/(2\pi)=4/97$]{\includegraphics[height=4cm]{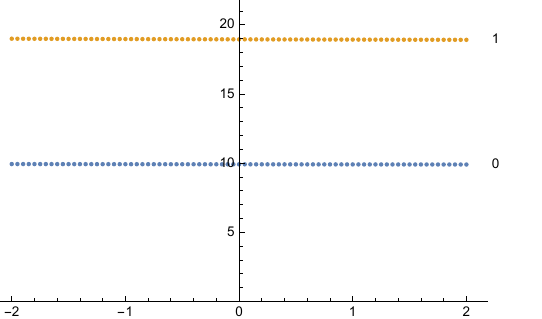}}%
	\hspace{2ex}%
	\subfloat[$\phi/(2\pi)=5/121$]{\includegraphics[height=4cm]{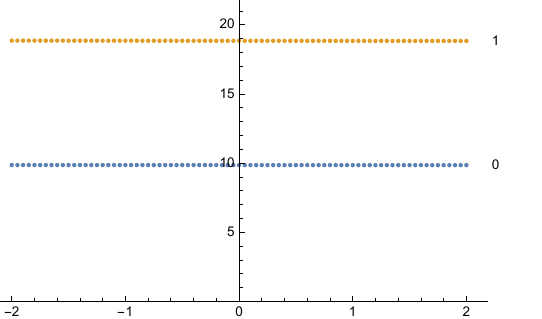}}%
	\hspace{2ex}%
	\subfloat[$\phi/(2\pi)=5/122$]{\includegraphics[height=4cm]{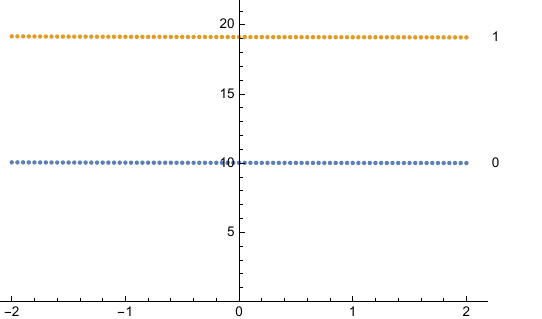}}%
	\hspace{2ex}%
	\subfloat[$\phi/(2\pi)=6/145$]{\includegraphics[height=4cm]{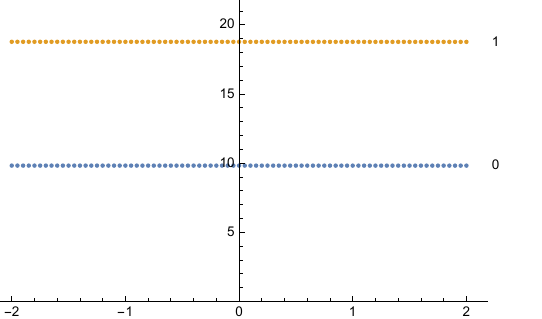}}%
	\caption{The orders of magnitude ($-\log_{10}(|*|)$, vertical axis)
		of the difference between the exact spectrum and the Borel
		resummation of full energy trans-seriesin the form of
		\eqref{eq:tensor} at Landau level 1 with varying $\Theta$
		(horizontal axis).  We include progressively contributions of
		increasing instanton orders $n=0,1$ from lower data points to
		higher data points.  The six plots are examples of (a)
		$P=2,Q=1\;\text{mod}\;P$ (a) $P=3,Q=1\;\text{mod}\;P$ (a)
		$P=4,Q=1\;\text{mod}\;P$ (a) $P=5,Q=1\;\text{mod}\;P$ (a)
		$P=5,Q=2\;\text{mod}\;P$ (a) $P=6,Q=1\;\text{mod}\;P$ for
		$\phi = 2\pi P/Q$.}
	\label{fig:spec-diff-PQ-lev1}
\end{figure}

For instance, we find the 1-instanton coefficient $w_{1,0}$ is one of
the $P$ solutions to
\begin{equation}
  \label{eq:u10}
  \Theta = \frac{1}{2}F_{Q/P}(2\pi u_{1,0},0,0)
  = : \frac{1}{2}F_{Q/P}(2\pi u_{1,0})
\end{equation}
where $F_{Q/P}(x)$ is the secular polynomial defined in
Sec.~\ref{sc:butterfly}.  Note that here the subscript is inverted
from $P/Q$ to $Q/P$, which may be related to the fractal structure of
the energy spectrum.  The secular polynomial has the property that
\begin{equation}
  F_{Q/P}(x) = F_{1-Q/P}(x).
\end{equation}
Some examples are
\begin{subequations}
  \begin{align}
    &F_1(x) = x,\\
    &F_{1/2}(x) = -4+x^2,\\
    &F_{1/3}(x) = -6x+x^3,\\
    &F_{1/4}(x) = 4-8x^2 + x^4,\\
    &F_{1/5}(x) = \frac{5}{2}(7-\sqrt{5})x-10x^3+x^5,\\
    &F_{2/5}(x) = \frac{5}{2}(7+\sqrt{5})x-10x^3+x^5,\\
    &F_{1/6}(x) = -4 + 24 x^2 -12 x^4 + x^6.
  \end{align}
\end{subequations}
Some numerical evidence of \eqref{eq:u10} are provided in
Fig.~\ref{fig:spec-diff-PQ-lev0} and Fig.~\ref{fig:spec-diff-PQ-lev1}.

It's quite clear from the right figure of Fig. (\ref{fig:splitting bands}) that the entire structure of Hofstadter's butterfly reemerges within each primary Landau level. From our preliminary analysis on higher instanton corrections for splitting bands, the higher trans-series coefficients are more complicated to be determined numerically and they are not naive generalizations of trans-series coefficients for $P=1$ case.

\subsection{Evidence for exact Rammal-Wilkinson formula}
\label{sc:EeRW}
Another approach to characterize these splitting bands is identifying the proper expansion base point of these bands such that the secondary Landau levels become primary Landau levels in this new expansion scenario. In \cite{wilkinson:1984example,rammal1990AnAS}, they discovered a
formula for perturbative energy expansion near arbitrary rational
values. However, due to limitations of technology at that age, their
computation of the perturbative energy from the quantization condition
were quite primitive and incomplete, i.e, the perturbative expansion
of Rammal-Wilkinson formula was calculated only to the first order and
their consideration of non-perturbative instanton corrections are
hand-waving without any determination of instanton action or
prefactor.

In order to perform similar numerical analysis for bands near certain
rational values as in \cite{Duan:2018dvj} and \cite{Hatsuda:2017dwx},
we need to introduce the concept of almost canonical continued
fraction. For a given non-negative rational number $\alpha$, it can always be
expressed as
\begin{equation}
  \alpha=n_0+\frac{1}{n_1+\frac{1}{n_2+\cdots+\frac{1}{n_l}}}
\end{equation}
and be denoted as $[n_0,n_1,n_2,\cdots,n_l]$. This is the canonical
continued fraction. Due to the symmetry property of the butterfly, we
can restrict our attention to real numbers that satisfy
$0\leq r\leq 1/2$ and $n_0$ can be set to 0. The almost canonical
continued fraction representation of a real number can be achieved by
allowing $n_i$ to be negative and requiring
\begin{equation}
  |n_i|\geq 2.
\end{equation} 
The representation of a rational number written in almost canonical
representation should be unique since we can always start with a
canonical continued fraction and rewrite
\begin{align}
  \begin{split}
    &[0,n_1,n_2,\cdots,n_{i-1},1,n_{i+1},n_{i+2},\cdots,n_l]\\
    \to &[0,n_1,n_2,\cdots,n_{i-1}+1,-(n_{i+1}+1),-n_{i+2},\cdots,-n_l]
  \end{split}
\end{align}
whenever we find $n_i=1$ in the sequence.

After representing the rational magnetic flux $\alpha = \phi/(2\pi)$
in the almost canonical fashion, $n_1$ is nothing but the number of
principal Landau levels, and $n_k$ is in general the number of
sub-levels at that nested layer assuming there's no merging of
subbands. For a rational magnetic flux
$\alpha = [0,n_1,n_2,\cdots,n_{l-1},n_l]$, we can regard it as a small
deviation from $\alpha_0 = [0,n_1,n_2,\cdots,n_{l-1}]$ and consider
the perturbative expansion of the energy from certain exact energy
value at $\alpha_0$, usually at an edge of energy bands.
In the examples shown below, we will focus on bands at $[0,n_1,n_2]$
expanded around $[0,n_1]=1/n_1$.

The simplest possible example to illustrate the expansion around
rational points other than zero would be considering
the base point $\alpha_0 = 1/2$, and taking the energy value at the top
edge of the first energy band $E_0 = 2\sqrt{2}$, solved from
\begin{equation}
  F_{1/2}(E,0,0) - 4 = 0.
\end{equation}
If we use
$\phi'=\phi-\phi_0$ with $\phi_0 = 2\pi\alpha_0 = \pi$ as our
expansion parameter, the perturbative series can be approximated
numerically from the exact spectra
\begin{align}
  \begin{split}
    E(\phi';N)= &2\sqrt{2} -\frac{(2 N+1)
      \phi^{\prime}}{\sqrt{2}}+\frac{\left(4 N^2+4
        N+3\right){\phi^{\prime}}^2}{8 \sqrt{2}}\\
    &+\frac{\left(8 N^3+12N^2+30N+13\right){\phi^{\prime}}^3}{96
      \sqrt{2}} +\mathcal{O}(\phi'^4).
  \end{split}
\end{align}
The leading instanton contribution to the bandwidths near $[0,2]$ can
also be approximated in reasonable precision,
\begin{equation}
  \mathrm{bw}_N(\phi')\simeq \frac{2^{2N+4}}{\sqrt{\pi}\,N!}
  \re^{-2C/\phi'}
  \bigg(1-\frac{3 N^2+9 N+5}{12} \phi^{\prime}-\frac{9 N^4+34 N^3+9 N^2-46 N-56}{288}\phi'^2+\mathcal{O}(\phi'^3)\bigg).
\end{equation}
We next consider the bands at $\alpha = [0,3,n_2]$ away from the base
point $\alpha_0 = 1/3$ expanded around the top band edge $E_0 =
\sqrt{3}+1$ solved from 
\begin{equation}
  F_{1/3}(E,0,0) - 4 = 0,
\end{equation}
as another case study.
If we use $\phi'=\phi-\phi_0$ with $\phi_0 = 2\pi\alpha_0 = 2\pi/3$ as
our expansion parameter, then
\begin{align}
	\begin{split}
		E(\phi';N)&=\sqrt{3}+1- \frac{3}{4}(\sqrt{3}-1)(2 N+1)|\phi'|+\left(\frac{\sqrt{3}}{2}-1\right)\phi'\\
		&+\frac{1}{32}\bigg(18(7 \sqrt{3}-11) N(N+1)+65 \sqrt{3}-95\bigg)\phi'^2\\
		&-\frac{3}{8}(5\sqrt{3}-8)(2N+1)\text{sgn}(\phi')\phi'^2+\mathcal{O}(\phi^3),
	\end{split}
\end{align}
and the bandwidth is approximated by
\begin{equation}
  \mathrm{bw}_N(\phi')\simeq \text{const.}\frac{2^{4N+\frac{9}{2}}}{\sqrt{\pi}\,N!
    \,3^{2N}}{\phi'}^{\frac{1}{2}-N}
  \re^{-8C/9\phi'}\mathcal{P}_1^{\text{inst}}({\phi'};N),
\end{equation}
where $\mathcal{P}_1^{\text{inst}}({\phi'};N)$ is a power series
starting from 1 that represents the instanton fluctuation.  If we use
instead $\tilde{\phi}$ as our expansion parameter, which is given by
\begin{equation}
  \label{eq:tphi}
  \phi=\frac{2\pi}{3-\frac{\tilde{\phi}}{2\pi}},
\end{equation}
the perturbative energy series is
\begin{equation}
  E(\tilde{\phi};N)=\sqrt{3}+1- \frac{1}{12}(\sqrt{3}-1)(2
  N+1)|\tilde{\phi}|+\left(\frac{\sqrt{3}-2}{18}\right)
  \tilde{\phi}+\mathcal{O}(\tilde{\phi}^2),
\end{equation}
and the energy bandwidth is approximated by
\begin{equation}\label{subbws}
  \mathrm{bw}_N(\tilde{\phi})\simeq
  \frac{2^{4N+\frac{9}{2}}}{\sqrt{\pi}\,N!}\tilde{\phi}^{\frac{1}{2}-N}
  \re^{-8C/\tilde{\phi}}\,\mathcal{P}_2^{\text{inst}}(\tilde{\phi};N),
\end{equation}
where $\mathcal{P}_2^{\text{inst}}$ is again a power series starting
from 1.  The prefactor and the instanton action of (\ref{subbws}) is
identical to the one appearing for bandwidths formula for $[0,n_1]$,
which suggests that \eqref{eq:tphi} perhaps is a more natural way of
performing the expansion. Extracting information of the instanton
fluctuation can help us learn about the quantum periods expanded near
the corresponding quantum conifold points. We wish to come back to
this problem in future works.

\section{Conclusion and discussion}
\label{sc:con}

In this paper, we begin to study the full energy trans-series for the
Harper-Hofstader model.  Using inspiration from the structure of
energy trans-series of 1d non-relativistic QM models obtained by the
exact WKB method, and the connection between the Harper-Hofstadter
model and the 5d SYM theory, we are able to write down a conjectural
full energy trans-series including instanton corrections at all levels
when the magnetic flux is $\phi = 2\pi/Q$, $Q\in\IN$, and we checked
our conjectural formula with very high numerical precisions, up to six
instanton levels.

One prominent feature of the full energy trans-series is that the
perturbative series only determines even instanton sectors via
resurgence but not the odd instanton sectors, which are in different
topological sectors, so that the strong resurgence program does not
hold.

When the magnetic flux is $\phi = 2\pi P/Q$ with $P>1$, although we
argue the resurgent structure of the perturbative series remains the
same, the coefficients of the full energy trans-series could be quite
different.  For instance, the coefficient of the 1-instanton sector is
given by roots of the secular equation with the inverted flux.  In
addition, we also made progress in the expansion of energy around a
rational value of magnetic flux instead of at the zero flux, including
both the perturbative energy series and the leading contribution to
energy bandwidth, extending the Rammal-Wilkinson formula.

There are many open problems following this work.  The energy spectrum
of the Harper-Hofstadter model is mesmerizing for the distinction
between rational and irrational values of the magnetic flux, and for
the self-similarity structure of the energy spectrum.
To understand the self-similarity structure of the energy spectrum, it
will be worthwhile to push further the calculation of the trans-series
coefficients for higher instanton levels when the magnetic flux is
$\phi = 2\pi P/Q$ with $P>1$.
One should also explore further the expansion of energy around a
non-zero rational value of magnetic flux.  One important line of
attack is to use the supersymmetric localization results of the Wilson
loop vev of 5d SYM \cite{Bullimore:2014upa,Bullimore:2014awa}, which
we argued to coincide with the energy of the Harper-Hofstadter model,
as it is more suitable for expansion around the rational value of magnetic
flux. \footnote{The Wilson loop vev, or equivalently the inverse quantum mirror map should be expanded around the rational value flux together with the identification of the perturbative quantum period as the perturbative quantization condition near the rational value flux in consideration.}
To understand the distinction between rational and irrational values
of the magnetic flux, it would be very beneficial to exploit the
relation between the Harper-Hofstadter model and the quantum group $\mc{U}_q(\mathfrak{sl}_2)$
\cite{Ikeda:2017ztr,Marra:2023gio} and quantum integrable models
\cite{Wiegmann:1994zz,Faddeev:1993uk}.
It would also be interesting to consider other lattices which are
related supersymmetric gauge theories or topological string
\cite{Hatsuda:2017zwn,Hatsuda:2017dwx,Hatsuda:2020ocr}.
It would also be interesting to perform a systematic exact WKB analysis on the Harper-Hofstadter model as previous studies on 4d SYM \cite{Kashani-Poor:2015pca,Sueishi:2021xti} as another line of attack.
We would like to return to these problems in the near future.

In an orthogonal direction, as we discussed in
Sec.~\ref{sc:summability}, we can calculate the Stokes constants of
the Harper-Hofstadter perturbative energy series from those of the
perturbative Wilson loop vevs in 5d SYM.  However, in this process, we
face the problem of choosing between using the Stokes constants in the strong
coupling regime or in the weak coupling regime from the 5d SYM.
A similar problem was already encountered in \cite{Marino:2024yme} where
one wished to reconstruct the Stokes constants of topological string
free energy in the conifold limit from the Stokes constants of
conventional topological string free energy.  The authors of
\cite{Marino:2024yme} proposed to use the Stokes constants in the
strong coupling regime, but could not provide an explanation.  Here we
find the same prescription is true, and we argue that the reason is
because the range of energy of the Harper-Hofstadter model is mapped
to the Coulomb modulus of the 5d SYM in the strong coupling regime.
We hope this argument can shed some light on the mystery in
\cite{Marino:2024yme}.





\section*{Acknowledgement}

We would like to thank Stavros Garoufalidis, Yasuyuki Hatsuda, Yunfeng
Jiang, Zhijin Li, Tadashi Okazaki, Maximilian Schwick, Ryo Suzuki,
Wenbin Yan for stimulating discussions, and thank Marcos Mari\~no for
carefully reading the manuscript.  We especially thank Stavros
Garoufalidis for sharing with us an unpublished manuscript of Armelle
Barelli, Jean Bellissard and Robert Fleckinger
\cite{Barelli:unpub}. We are especially grateful to Alexander van
Spaendonck and Marcel Vonk for clarifying some subtleties of tensor
factorization of trans-series. 
We also thank the workshop ``String Theory and Quantum Field Theory
2024'' hosted at Fudan University, where this work was initiated,
thank the workshop on ``Non-Perturbative and Enumerative Aspects in
Topological Theories from Geometric Engineering'' organised by IASM at
Zhejiang University, where part of the work was completed, and thank
the Lunch Seminar Series of Shing-Tung Yau Center at Southeast
University, the workshop ``Forum on Supersymmetry in Physics and
Mathematics'' organised by ICTP-AP, and the workshop ``Resurgence
Theory in Mathematical Physics'' organised by Chern Institute of
Mathematics at Nankai University, where the results of this work were
presented. J.G. is supported by the startup funding No. 4007022316 and
4007022411 of the Southeast University, and the National Natural
Science Foundation of China (General Program) funding No. 12375062.


\bibliographystyle{JHEP}
\bibliography{hofstadter}

\end{document}